\documentclass[showpacs,showkeys,
preprintnumbers,eqsecnum,prd,
longbibliography,
nofootinbib,amsfonts,amsmath,amssymb
]{revtex4-1}

\usepackage[english]{babel}

\usepackage[usenames]{color}
\usepackage{graphicx}
\usepackage{latexsym}
\usepackage{amssymb}
\usepackage{slashbox}

\usepackage{hyperref}
\usepackage{mathrsfs}
\usepackage{multirow}

\definecolor{darkgreen}{rgb}{0,0.5,0}

\hypersetup{
    bookmarks=true,         
    unicode=false,          
    pdftoolbar=true,        
    pdfmenubar=true,        
    pdffitwindow=false,     
    pdfstartview={FitH},    
    pdftitle={My title},    
    pdfauthor={Author},     
    pdfsubject={Subject},   
    pdfcreator={Creator},   
    pdfproducer={Producer}, 
    pdfkeywords={keyword1} {key2} {key3}, 
    pdfnewwindow=true,      
    colorlinks=true,       
    linkcolor=red,          
    citecolor=cyan,        
    filecolor=magenta,      
    urlcolor=darkgreen,           
    linktocpage=true
}

\expandafter\let\csname equation*\endcsname=\relax
\expandafter\let\csname endequation*\endcsname=\relax

\allowdisplaybreaks

\def\beq{\begin{equation}}
\def\eeq{\end{equation}}
\def\rmd{{\rm d}}

\begin{document}

\title{Dynamics of extended bodies in a Kerr spacetime with spin-induced
  quadrupole tensor}

\author{Donato Bini} \affiliation{ Istituto per le Applicazioni del Calcolo
  ``M. Picone,'' CNR, I-00185 Rome, Italy }

\author{Guillaume Faye}
\affiliation{$\mathcal{G}\mathbb{R}\varepsilon{\mathbb{C}}\mathcal{O}$,
  Institut d'Astrophysique de Paris, UMR 7095 CNRS, Sorbonne Universit{\'e}s,
  UPMC Univ Paris 06, F-75014 Paris, France}

\author{Andrea Geralico}
  \affiliation{Istituto per le Applicazioni del Calcolo ``M. Picone,'' CNR,
    I-00185 Rome, Italy}

\date{\today}

\begin{abstract}

  The features of equatorial motion of an extended body in Kerr spacetime are
  investigated in the framework of the Mathisson-Papapetrou-Dixon model. The
  body is assumed to stay at quasi-equilibrium and respond instantly to
  external perturbations. Besides the mass, it is completely determined by
  its spin, the multipolar expansion being truncated at the quadrupole order,
  with a spin-induced quadrupole tensor. The study of the radial effective
  potential allows to analytically determine the ISCO shift due to spin and
  the associated frequency of the last circular orbit.

\end{abstract}


\maketitle


\section{Introduction}

The description of the gravitational interaction between the constituents of a
binary system in the general theory of relativity requires taking into due
account their internal structure. The orbital dynamics of two bound compact
objects is tackled in the literature by a plenty of different methods
resorting to various approximation schemes. Analytic approaches include
notably the post-Newtonian approximation~\cite{Bliving}, possibly implemented
using effective field theory techniques~\cite{Goldberger07}, as well as the
gravitational self-force corrections to geodesic motion~\cite{Poisson:2011},
which can both be combined efficiently by means of the ``effective-one-body''
approach~\cite{Buonanno:1998gg,Damour14}.

When the mass of one body is much smaller than the other, the problem boils
down to studying the dynamics of an extended body in a fixed background field,
generated by the heavier mass. In this approximation, a self-consistent model
describing the evolution of both linear and angular momenta for pole-dipole
sources was developed by Mathisson~\cite{math37},
Papapetrou~\cite{papa51,corpapa51}, Pirani~\cite{pir56},
Tulczyjew~\cite{tulc59}, and later generalized to bodies endowed with higher
multipoles by Dixon~\cite{dixon64,dixon69,dixon70,dixon73,dixon74}. The
Mathisson-Papapetrou-Dixon (MPD) model accounts for the motion, on a fixed
background, of a point-size test object with internal degrees of freedom, in
the absence of significant gravitational back reaction.

The main astrophysical situation for which a full relativistic treatment is
needed occurs when the object --- assumed to be compact to avoid tidal
disruption --- experiences the strong field produced by a nearby black hole.
In that case, the MPD approach may be used to investigate the evolution of the
system, but the parameters of the model must be regarded as effective
ones~\cite{Harte12}. Although self-force effects are not negligible on
larger-than-orbital time scales, they only yield higher-order corrections. On
the other hand, combined with dimensional regularization, the MPD model is
appropriate to describe the dynamics of (self-gravitating) compact binaries
including finite-size effects, in the post-Newtonian
framework~\cite{OTO98,FBB06,BFMP15}. It yields results that are dynamically
equivalent to those derived from suitable effective
actions~\cite{porto06,porto06prl,Barausse:2009aa}. The ``skeleton''
stress-energy tensor encoding the MPD evolution is known at the quadrupolar
level~\cite{SP10}. Its octupolar contributions have also been obtained
recently~\cite{Marsat15} assuming an effective, Bailey \& Israel type,
Lagrangian~\cite{BI75}. Those corresponding explicit expressions have been
used to build accurate theoretical templates for the signal of gravitational
radiation emitted by those
sources~\cite{Marsat:2013caa,BFMP15,Marsat15,BFH12}, in the context of the
data analysis of gravitational-wave observatories, such as the advanced
Virgo~\cite{virgo} and LIGO~\cite{ligo} detectors, the future cryogenic
interferometer KAGRA~\cite{kagra} or, possibly, the space-based observatory
eLISA~\cite{Whitepaper}, a candidate for the future L3 mission of the European
Space Agency. Most of post-Newtonian expressions can be checked by comparing
them to the test-body counterparts, in the extreme mass-ratio limit.

In this paper we study the dynamics of an extended body endowed with both spin
and quadrupole moment in a Kerr spacetime using the MPD model. The motion is
assumed to be confined on the equatorial plane, the spin vector of the body
being aligned with the axis of rotation of the central object. In previous
works, we have discussed the effects on the dynamics of a general quadrupole
tensor in both Schwarzschild and Kerr spacetimes
\cite{quadrupschw,quadrupkerr1,quadrupkerrnum}. Here, we consider more
specifically the case of a spin-induced quadrupole tensor. We assume that the
object reaches thermodynamic equilibrium in its proper frame on time
scales that are very short compared with the orbital period and neglect the
tidal deformations. Its internal state thus depends adiabatically on the mass
and the (instantaneous) spin. Using effective field theory arguments, it is
then straightforward to check that the body quadrupole is actually quadratic
in the spin. This situation was described in details by Steinhoff \&
Puetzfeld~\cite{steinhoff}, who developed a very general framework to include
quadratic in spin corrections as well as tidal interactions in the MPD scheme,
with special attention to the study of the binding energy of the system as
obtained from the analysis of the associated effective potential. Later,
Hinderer et al.~\cite{hinderer} performed an analysis of the corresponding
dynamics, in order to compare the periastron advance and precession
frequencies with those of a different approach, but restricted themselves to a
very special (although physically motivated) choice of the quadrupole tensor,
leading to great simplifications in the analytic treatment.

Here, we shall assume the same form of the quadrupole tensor as in
Ref.~\cite{steinhoff}, but neglect quadrupolar tides, i.e., our quadrupole
tensor is of the electric-type only and is proportional to the trace-free part
of the square of the spin tensor by a constant parameter, whose numerical
value is a property of the body under consideration. For neutron stars such a
quantity depends on the equation of state~\cite{poisson}, while it is exactly
1 for black holes. We keep it a free parameter of the model that can affect
associated observables, like the energy and the angular momentum, which we
computed explicitly and compared with the results of Ref.~\cite{hinderer}, or
the Innermost Stable Circular Orbit (ISCO) and its frequency, discussed here in
details. We achieve a fully analytic treatment of the MPD equations in a
``perturbative'' scheme, obtaining corrections to geodesic motion up to the
second order in spin.

Throughout this work we use geometrical units, setting the Newton constant $G$
and the speed of light $c$ to 1. Tensors are represented either in abstract
notation or in index notation combined with the Einstein convention, depending
on the context. Greek indices refer to spacetime coordinates and vary from 0
to 3, i.e., $\mu=0,1,2,3$, whereas Latin indices, ranging from 1 to 3, label
space coordinates. The notation $\partial_\mu$ stands indifferently for the
partial derivative with respect to the $\mu$th coordinate $x^\mu$ or for the
coordinate basis vector associated to $x^{\mu}$, while $\rmd$ denotes the
exterior derivative. The spacetime metric $g_{\mu\nu}$, taken to have
signature $(-,+,+,+)$, defines a unique Levi-Civita covariant derivative,
$\nabla_\mu$ and an associated Riemann curvature $R^{\mu}{}_{\nu\alpha\beta}$,
with the convention that $R^{\mu}{}_{\nu\alpha\beta} v^\nu =(\nabla_\alpha
\nabla_\beta - \nabla_\beta \nabla_\alpha) v^\mu$ for any vector field
$v^\mu$. Symmetrization of a tensor $T$ over a set indices is indicated by
round brackets enclosing them: $T^{(\mu\nu)} = (T^{\mu\nu} + T^{\nu\mu})/2$.
Instead, for index antisymmetrization, square brackets are used: $T^{[\mu\nu]}
= (T^{\mu\nu} - T^{\nu\mu})/2$.


\section{MPD description of extended bodies} \label{MPD}

In the quadrupole approximation, the MPD equations read
\begin{align}
\label{papcoreqs1}
\frac{{\rm D}P^{\mu}}{\rmd \tau} & = 
- \frac12 \, R^\mu{}_{\nu \alpha \beta} \, U^\nu \, S^{\alpha \beta}
-\frac16 \, \, J^{\alpha \beta \gamma \delta} \, \nabla^\mu R_{\alpha \beta \gamma \delta}
\nonumber\\
& \equiv  F^\mu_{\rm (spin)} + F^\mu_{\rm (quad)} \,,
\\
\label{papcoreqs2}
\frac{{\rm D}S^{\mu\nu}}{\rmd \tau} & =  
2 \, P^{[\mu}U^{\nu]}+
\frac43 \, J^{\alpha \beta \gamma [\mu}R^{\nu]}{}_{\gamma \alpha \beta}
\nonumber\\
&\equiv  D^{\mu \nu}_{\rm (spin)} + D^{\mu \nu}_{\rm (quad)} \,,
\end{align}
where $P^{\mu}\equiv m u^\mu$ (with $u \cdot u \equiv u^\mu u_\mu = -1$) is
the total 4-momentum of the body with mass $m$ and direction $u^\mu$, $S^{\mu
  \nu}$ is the (antisymmetric) spin tensor, $J^{\alpha\beta\gamma\delta}$ is
the quadrupole tensor, and $U^\mu=\rmd z^\mu/\rmd\tau$ is the timelike unit
tangent vector --- or 4-velocity --- of the body ``reference'' line,
parametrized by the proper time $\tau$ (with parametric equations
$x^\mu=z^\mu(\tau)$), used to make the multipole reduction.

In order to ensure that the model is mathematically self-consistent, the
reference point in the object should be specified by imposing some additional
conditions. Here we shall take the Tulczyjew conditions~\cite{tulc59,dixon64},
\beq
\label{tulczconds}
S^{\mu\nu}u{}_\nu=0\,. 
\eeq
With this choice, the spin tensor can be fully represented by a spatial vector
(with respect to $u$),
\beq
S(u)^\alpha=\frac12
\eta(u)^\alpha{}_{\beta\gamma}S^{\beta\gamma}=[{}^{*_{(u)}}S]^\alpha\,,
\eeq 
where $\eta(u)_{\alpha\beta\gamma}=\eta_{\mu\alpha\beta\gamma}u^\mu$ is the
spatial unit volume 3-form (with respect to $u$) built from the 4-volume form
$\eta_{\alpha\beta\gamma\delta}=\sqrt{-g}\, \epsilon_{\alpha\beta\gamma\delta}$,
with $\epsilon_{\alpha\beta\gamma\delta}$ ($\epsilon_{0123}=1$) being the
Levi-Civita alternating symbol and $g$ the determinant of the metric in a
generic coordinate grid. Using a fairly standard convention, hereafter we
denote the spacetime dual of a tensor $S$ (such that $^*S^{\alpha\beta} =
\eta^{\alpha\beta}_{\phantom{\alpha\beta}\gamma\delta} S^{\gamma\delta}/2$) by
$^*S$, whereas the spatial dual of a spatial tensor $S$ with respect to $u$ is
represented by $^{*_{(u)}}S$. It is also useful to introduce the signed
magnitude $s$ of the spin vector, which is not constant in general along the
trajectory of the extended body:
\beq
\label{sinv}
s^2=S(u)^\beta S(u)_\beta = \frac12 S_{\mu\nu}S^{\mu\nu}
=-\frac12{\rm Tr}[S^2]\,,
\eeq
with ${\rm Tr} [S^2]= -S_{\mu\nu}S^{\mu\nu}$.

\subsection{Spin-induced quadrupole tensor}

The $1+3$ decomposition of the quadrupole tensor $J$ and its general
properties are briefly reviewed in Appendix~\ref{appJ}. We will consider here
the physically relevant case where it is completely determined by the
instantaneous spin structure of the body (see, e.g., Refs.
\cite{steinhoff,steinhoff2}). More specifically, we shall let the quadrupole
tensor have the form
\beq
\label{Jspininduced}
J^{\alpha\beta\gamma\delta}=
4{u}^{[\alpha}\widetilde{\mathcal X}(u)^{\beta][\gamma}{u}^{\delta]}\,, 
\eeq
with
\beq
\widetilde{\mathcal X}(u)=\frac34\frac{C_Q}{m}[S^2]^{\rm STF}\,,
\eeq
where $C_Q$ is a ``polarizability'' constant and $[S^2]^{\rm STF}$ denotes the
trace-free part of $S^{\alpha\mu}S_{\mu}{}^{\beta}$, i.e., in terms of both
the spin vector and the associated spin invariant,
\begin{align}
[S^2]^{\rm STF}{}^{\alpha\beta}=S^{\alpha\mu}S_{\mu}{}^{\beta}
-\frac13P(u)^{\alpha\beta}S_{\rho\sigma}S^{\sigma\rho}
=S(u)^{\alpha}S(u)^{\beta}-\frac13s^2P(u)^{\alpha\beta}\,.
\end{align}
The values of $C_Q$ associated with compact objects are given, e.g., in
Ref.~\cite{steinhoff3}. The normalization is such that $C_Q=1$ in the case of
a black hole~\cite{thorne}, whereas for neutron stars $C_Q$ depends on the
equation of state and varies roughly between 4 and 8~\cite{poisson}.

For the spin-induced quadrupole tensor~\eqref{Jspininduced}, the link between
$P_\mu$ and $U_\mu$ takes a particularly simple form if cubic-in-spin
corrections are neglected. Indeed, contraction of Eq.~\eqref{papcoreqs2} with
$P_\nu$ shows that, apart from corrections of order $J=\mathcal{O}(S^2)$, the
difference $(P \cdot U) P^\mu - P^2 U^\mu$ is precisely $P_\nu {\rm D}S^{\mu\nu}
/\rmd \tau =-S^{\mu\nu} {\rm D} P_\nu/\rmd \tau $, which can be seen to be of
order $\mathcal{O}(S^2)$ from Eq.~\eqref{papcoreqs1}. In the end, $P^\mu$ is
approximately proportional to $U^\mu$ and, as an important consequence, the
right-hand side of the precession equation~\eqref{papcoreqs2} is at least
quadratic in the spin. As, on the other hand, the right-hand side of the
precession equation~\eqref{papcoreqs2} is at least linear in the spin, we
conclude that the time differentiation of our kinematical variables
actually multiply any combination of them by a factor $\mathcal{O}(S)$.

In Ref.~\cite{hinderer}, the MPD description of test bodies endowed with a
spin-induced quadrupolar structure is used to check the consistency of the
computation of the periastron advance for a binary system with the
effective-one-body formalism in the extreme mass-ratio limit, including terms
that are quadratic in the spin. The quadrupole tensor is assumed there to be
that of a black hole and to take the form \eqref{Jspininduced} with $C_Q=1$,
so that we must recover the results of Ref.~\cite{hinderer} for this
particular value.

\subsection{Simplified form of the MPD equations}
\label{simplifiedMPD}

The MPD equations can be written in a more convenient form at quadratic order
in the spin. First, the quadrupolar contribution $F^\mu_{\rm (quad)}$ in
Eq.~\eqref{papcoreqs1} splits into a parallel and a perpendicular part with
respect to the direction $U^\mu$:
\begin{align}
F^\mu_{\rm (quad)}
=  -\frac16 \, \, J^{\alpha \beta \gamma \delta} \, [P(U)]^{\mu\nu}\nabla_\nu
R_{\alpha \beta \gamma \delta} + \frac16 \, \, J^{\alpha \beta \gamma \delta}
\,  U^\mu \frac{\rm D}{\rmd \tau} R_{\alpha \beta \gamma \delta} \, .
\end{align}
Next, neglecting remainders that are cubic in the spin, the quadrupole
$J^{\alpha \beta \gamma \delta}$ and the velocity $U^\mu$ in the last term may
be moved under the operator $\mathrm{D}/\rmd \tau$, since their covariant time
differentiation would actually produce terms smaller than the original ones by
a factor $\mathcal{O}(S)$, as explained in the previous subsection. The
equations of motion then become:
\begin{align}
\frac{{\rm D}P^{\mu}}{\rmd \tau} = 
- \frac12 \, R^\mu{}_{\nu \alpha \beta} \, U^\nu \, S^{\alpha \beta}
-\frac16 \, \, J^{\alpha \beta \gamma \delta} \, [P(U)]^{\mu\nu}\nabla_\nu
R_{\alpha \beta \gamma \delta} + \frac{\rm D}{\rmd \tau} \left(
m_J
U^\mu \right) \, , 
\end{align}
where we have posed
\beq
m_J=\frac16 \, \, J^{\alpha \beta \gamma \delta} 
R_{\alpha \beta \gamma \delta}\,.
\eeq
Thus, defining a modified linear momentum $p^\mu= P^{\mu} - m_JU^\mu$
effectively changes the quadrupolar force $F^\mu_{\rm (quad)}$ into its
projection $[P(U)]^{\mu}_{~\nu}F^\nu_{\rm (quad)}$ orthogonal to $U^\mu$,
while the spin force $F^\mu_{\rm (spin)}$ is orthogonal to $U^\mu$. The
precession equations are unaffected. More explicitly, Eqs.~\eqref{papcoreqs1}
and~\eqref{papcoreqs2} expressed in terms of $p^\mu$ become:
\begin{subequations}
\begin{align}
\frac{{\rm D}p^{\mu}}{\rmd \tau} &=
- \frac12 \, R^\mu{}_{\nu \alpha \beta} \, U^\nu \, S^{\alpha \beta}
-\frac16 \, \, J^{\alpha \beta \gamma \delta} \,  \nabla^{\perp}
R_{\alpha \beta \gamma \delta} + \mathcal{O}(S^3)\, ,  \label{papcoreqsmodif1}
\\
\frac{{\rm D}S^{\mu\nu}}{\rmd \tau}  &=
2 \, p^{[\mu}U^{\nu]}+
\frac43 \, J^{\alpha \beta \gamma [\mu}R^{\nu]}{}_{\gamma \alpha \beta} +
\mathcal{O}(S^3) \, , \nonumber
\end{align}
\end{subequations}
with $\nabla^{\perp}\equiv[P(U)]^{\mu\nu}\nabla_\nu$. Now, if we multiply the
equations of motion~\eqref{papcoreqsmodif1} by $p_\mu \propto U_\mu +
\mathcal{O}(S^2)$, we see that, at our accuracy level, the first term on the
right-hand side vanishes by virtue of the Riemann-tensor symmetries whereas
the second one is zero due to the contraction of $U_\mu$ with
$[P(U)]^{\mu\nu}$. We conclude that the effective mass $(-p_\mu p^\mu)^{1/2}$
is conserved, modulo cubic spin corrections (see also Appendix B). It may be
regarded as the ``bare'' mass of the extended body, $m_0$, at the quadratic
order in the spin, so that its mass $m$ to order $\mathcal{O}(S^2)$ is given
by
\beq
\label{mJdef}
m=m_0+m_J+\mathcal{O}(S^3)\,.
\eeq
Finally, the MPD equations \eqref{papcoreqs1}--\eqref{tulczconds} imply
that the unit vectors $U$ and $u$ are related by
\beq
\label{reluUgen}
u^\mu=U^\mu+\frac1{m_0}D^{\mu \nu}_{\rm (quad)}U_\nu
+\frac1{m_0^2}S^{\mu\nu} F_{\rm (spin)}{}_{\nu}+\mathcal{O}(S^3)\,.
\eeq 

\subsection{Conserved quantities}

In stationary and axisymmetric spacetimes endowed with Killing symmetries,
the energy $E$ and the total angular momentum $J$ are conserved quantities
along the motion, associated with the timelike Killing vector $\xi=\partial_t$
and the azimuthal Killing vector $\eta=\partial_\phi$, respectively. 
They are given by
\begin{align}
\label{totalenergy}
E&=-\xi_\alpha P^\alpha +\frac12 S^{\alpha\beta}F^{(t)}_{\alpha\beta}\,,\nonumber\\
J&=\eta_\alpha P^\alpha -\frac12 S^{\alpha\beta}F^{(\phi)}_{\alpha\beta}\,,
\end{align}
where
\beq
F^{(t)}_{\alpha\beta}=\nabla_\beta \xi_\alpha=g_{t[\alpha,\beta]}\,, \quad
F^{(\phi)}_{\alpha\beta}=\nabla_\beta \eta_\alpha=g_{\phi[\alpha,\beta]}\,, 
\eeq
are the Papapetrou fields associated with the Killing vectors. Note that $E$
and $J$ as defined above are conserved to all multipole orders in spite of the
higher multipolar structure of the body, which is entirely encoded in
$P$~\cite{ehlers77}.

The conserved quantities~\eqref{totalenergy} for a purely dipolar particle in
a Kerr spacetime have been computed, e.g., in Refs.~\cite{maeda1,maeda2}. The
expressions given there are general enough to account for all higher-order
spin corrections but those coming from the spin-induced multipole moments
(i.e., are exact when the quadrupole and higher multipole moments vanish).
This means in practice that our results should reduce to those of
Refs.~\cite{maeda1,maeda2} for $C_Q=0$.


\section{Motion in a Kerr spacetime}

The Kerr metric in standard Boyer-Lindquist coordinates $(t,r,\theta,\phi)$
reads
\begin{align}
\rmd s^2 &= -\left(1-\frac{2Mr}{\Sigma}\right)\rmd t^2 
-\frac{4aMr}{\Sigma}\sin^2\theta\,\rmd t\, \rmd\phi+ \frac{\Sigma}{\Delta}\rmd
r^2+\Sigma\,\rmd \theta^2+\frac{(r^2+a^2)^2
-\Delta a^2\sin^2\theta}{\Sigma}\sin^2 \theta \,\rmd \phi^2\,,
\end{align}
with $\Delta=r^2-2Mr+a^2$ and $\Sigma=r^2+a^2\cos^2\theta$. Here, $a\geq0$ and
$M \ge a$ denote the specific angular momentum and the total mass of the
spacetime solution, respectively. The event and inner horizons are located at
$r_\pm=M\pm\sqrt{M^2-a^2}$.

Let us introduce the zero angular-momentum observer (ZAMO) family of fiducial
observers, with 4-velocity
\beq
\label{n}
n=N^{-1}(\partial_t-N^{\phi}\partial_\phi)
\eeq
orthogonal to the hypersurfaces of constant $t$, where $N=(-g^{tt})^{-1/2}$
and $N^{\phi}=g_{t\phi}/g_{\phi\phi}$ are the lapse and shift functions,
respectively. A suitable spatial orthonormal frame adapted to ZAMOs is given
by
\beq
\label{ZAMO-frame}
e_{\hat t}=n\,, \quad 
e_{\hat r}=\frac1{\sqrt{g_{rr}}}\partial_r\equiv\partial_{\hat r}\,,\quad
e_{\hat \theta}=
\frac1{\sqrt{g_{\theta \theta }}}
\partial_\theta\equiv\partial_{\hat\theta}\,, \quad
e_{\hat \phi}=\frac1{\sqrt{g_{\phi \phi }}}
\partial_\phi\equiv\partial_{\hat\phi}\,,
\eeq
with dual 
\beq
\label{ZAMO-frame-dual}
\omega^{{\hat t}}=N\rmd t\,, \quad 
\omega^{{\hat r}}=\sqrt{g_{rr}}\,\rmd r\,, \quad
\omega^{{\hat \theta}}= \sqrt{g_{\theta \theta }} \,\rmd \theta\,, \quad
\omega^{{\hat \phi}}=\sqrt{g_{\phi \phi }}(\rmd \phi+N^{\phi}\rmd t)\,.
\eeq

The ZAMOs are subject to the acceleration $a(n)=\nabla_n n$. They are locally
non-rotating, in the sense that their vorticity vector $\omega(n)^\alpha$
vanishes due to their surface-orthogonal character, but they have a nonzero
trace-free expansion tensor $\theta(n)_{\alpha\beta}\equiv P(n)^\mu_{~\alpha}
P(n)^\nu_{~\beta} \nabla_{(\mu} n_{\nu)}$; the latter, in turn,
can be completely described by an expansion vector 
$\theta_{\hat \phi}(n)^\alpha=\theta(n)^\alpha{}_\beta\,{e_{\hat\phi}}^\beta$,
such that
\beq
\label{exp_zamo}
\theta(n) = e_{\hat\phi}\otimes\theta_{\hat\phi}(n)
+\theta_{\hat\phi}(n)\otimes e_{\hat\phi}
\,,
\eeq
where $\otimes$ represents the tensor product. The nonzero ZAMO kinematical
quantities (i.e., acceleration and expansion) all belong to the $r$-$\theta$
2-plane of the tangent space \cite{mfg,idcf1,idcf2,bjdf}, with
\begin{align}
\label{accexp}
a(n) & = a(n)^{\hat r} e_{\hat r} + a(n)^{\hat\theta} e_{\hat\theta}
\equiv\partial_{\hat r}(\ln N) e_{\hat r} + \partial_{\hat\theta}(\ln N) e_{\hat\theta}
\,,
\nonumber\\
\theta_{\hat\phi}(n) & = \theta_{\hat\phi}(n)^{\hat r}e_{\hat r} 
+ \theta_{\hat\phi}(n)^{\hat\theta}e_{\hat \theta} 
\equiv-\frac{\sqrt{g_{\phi\phi}}}{2N}\,(\partial_{\hat r} N^\phi e_{\hat r} 
+ \partial_{\hat\theta} N^\phi e_{\hat \theta})
\,.
\end{align}
It is also useful to introduce the curvature vectors
associated with the diagonal metric coefficients,
\begin{align}
\label{k_lie}
\kappa(x^i,n)
& = \kappa(x^i,n)^{\hat r} e_{\hat r} + \kappa(x^i,n)^{\hat\theta} 
e_{\hat\theta}\nonumber\\
 & \equiv -[\partial_{\hat r}(\ln \sqrt{g_{ii}}) e_{\hat r} 
+ \partial_{\hat\theta}(\ln \sqrt{g_{ii}})e_{\hat\theta}]
\,.
\end{align}
We shall use the notation $\kappa(\phi,n)^{\hat r}\equiv k_{\rm (Lie)}$ for
the Lie relative curvature \cite{idcf1,idcf2}, largely adopted in the
literature, and limit our analysis to the equatorial plane
($\theta=\pi/2$) of the Kerr solution, where
\beq
N=\left[\frac{r\Delta}{r^3+a^2r+2a^2M}\right]^{1/2}\,,\qquad
N^\phi=-\frac{2aM}{r^3+a^2r+2a^2M}\,,
\eeq
and $\Delta=N^2g_{\phi\phi}$.  The ZAMO kinematical
quantities as well as the nonvanishing frame components of the Riemann tensor
are listed in Appendix \ref{appzamos}.

Let us now consider a test body rotating in the equatorial plane around the
central source. Its $4$-velocity $U$ may be written in terms of the velocity
$\nu(U,n)=\nu^{\hat r} e_{\hat r}+\nu^{\hat \phi} e_{\hat \phi}$ relative to
the ZAMOs, with associated Lorentz factor $\gamma(U,n)$, as
\beq
\label{polarnu}
U=\gamma(U,n) [n+ \nu(U,n)]\,, \qquad
\gamma(U,n)=\left(1-||\nu(U,n)||^2\right)^{-1/2}\,,
\eeq
The parametric equations of the orbit are solutions of the
evolution equations $U=\rmd x^\alpha/\rmd\tau$, i.e.,
\begin{align}
\label{Uequatgen}
\frac{\rmd t}{\rmd \tau}&=\frac{\gamma}{N}\,,\qquad
\frac{\rmd r}{\rmd \tau} =\frac{\gamma\nu^{\hat r}}{\sqrt{g_{rr}}}\,,\nonumber\\
\frac{\rmd \phi}{\rmd \tau}&=
\frac{\gamma}{\sqrt{g_{\phi\phi}}}\left(\nu^{\hat \phi}
-\frac{\sqrt{g_{\phi\phi}}N^\phi}{N}\right)\,, 
\end{align}
where the abbreviated notation $\gamma(U,n)\equiv\gamma$ and $\nu^{\hat
  a}\equiv\nu(U,n)^{\hat a}$ has been used. For equatorial orbits,
a convenient parametrization can be $r$ itself instead of the proper time
$\tau$.

A case of particular importance is that of uniform, circular equatorial
motion. The unit tangent vector, $U$, may then be parametrized either by the
(constant) angular velocity with respect to infinity $\zeta$ or equivalently
by the (constant) linear velocity $\nu$ with respect to the ZAMOs, i.e.,
\beq
\label{orbita}
U=\Gamma [\partial_t +\zeta \partial_\phi ]=
\gamma [e_{\hat t} +\nu e_{\hat \phi}]\,, \qquad \gamma=(1-\nu^2)^{-1/2}\,,
\eeq
with
\beq
\label{orbitaGz}
\Gamma =\left[ N^2-g_{\phi\phi}(\zeta+N^{\phi})^2 \right]^{-1/2} = 
\frac{\gamma}{N}\,,\qquad
\zeta=-N^{\phi}+\frac{N}{\sqrt{g_{\phi\phi}}}\,\nu \,.
\eeq
The parametric equations of the orbit reduce to
\beq
\label{casocircorb}
t=t_0+\Gamma \tau\,,\quad 
r=r_0\,,\quad 
\theta=\frac{\pi}{2}\,,\quad
\phi=\phi_0+\Omega \tau\,,
\eeq
where $\Omega =\Gamma\zeta$ is the proper time orbital angular velocity.

For timelike circular geodesics on the equatorial plane, the expressions of
the angular and linear velocities do depend on whether the orbits are
co-rotating $(+)$ or counter-rotating $(-)$. They read
\beq
\label{geocirc}
\zeta_{\pm}=\pm\frac{\zeta_K}{1\pm a\zeta_K}\,,\qquad
\nu_\pm=\frac{r^2\zeta_\pm}{\sqrt{\Delta}}
\left(1+\frac{a^2}{r^2}\mp2a\zeta_K\right)\,,
\eeq
respectively, with $\zeta_K=\sqrt{M/r^3}$ denoting the Keplerian angular
velocity for a non-spinning, Schwarzschild black hole. The Lorentz factor of
the corresponding 4-velocity $U_\pm$ is found to be
\beq
\Gamma_\pm=\frac{\zeta_K}{|\zeta_\pm|}
\left(1-\frac{3M}{r}\pm 2a\zeta_K\right)^{-1/2}\,.
\eeq
In the static case, we can actually use the Schwarzschild values
$\zeta_\pm\to\pm\zeta_K$ and $\nu_\pm\to\pm\nu_K$, with
$\nu_K=\sqrt{M/(r-2M)}$. It is convenient [see Eq.~\eqref{relUu}] to
introduce a spacelike unit vector $\bar U_{\pm}$ that is orthogonal to
$U_{\pm}$ within the Killing 2-plane, by defining
\beq
\label{barUgeo}
\bar U_{\pm}=\bar \Gamma_\pm [\partial_t +\bar \zeta_{\pm} \partial_\phi ]
=\pm\gamma_{\pm} [\nu_{\pm} e_{\hat t} +e_{\hat \phi}]\,,
\eeq
in terms of the parameters
\begin{align}
\bar\Gamma_\pm&=\Gamma_{\pm}|\nu_{\pm}|
=|\Omega_\pm|\frac{r^2}{\sqrt{\Delta}} 
\left(1+\frac{a^2}{r^2}\mp2a\zeta_K\right)\,,
\nonumber\\
\bar\zeta_\pm&=-N^{\phi}+\frac{N}{\sqrt{g_{\phi\phi}}}\, \frac{1}{\nu_{\pm}}
=\pm\frac{r\zeta_K}{M}\frac{1-{2M}/{r}\pm a\zeta_K}{1+{a^2}/{r^2}\mp2a\zeta_K}\,,
\end{align}
where $\Omega_\pm=\Gamma_\pm\zeta_\pm$ and where the $\pm$ signs correlate with
those of $U_\pm$. Note that $\bar\zeta_\pm={\tilde E_\pm}/{\tilde L_\pm}$ is
the ratio between the energy $\tilde E_\pm$ and the azimuthal angular momentum
$\tilde L_\pm$ per unit mass of the particle. Those two quantities are
expressed as
\begin{align}
\label{EandLgeocirc}
\tilde E_\pm&=N\gamma_\pm\left(1+\frac{2aM}{r\sqrt{\Delta}}\nu_\pm\right)
=\frac{|\Omega_\pm|}{\zeta_K}\left(1-\frac{2M}{r}\pm a\zeta_K\right)\,,
\nonumber\\
\tilde L_\pm&=\gamma_\pm\nu_\pm\sqrt{g_{\phi\phi}}
=\Omega_\pm r^2\left(1+\frac{a^2}{r^2}\mp2a\zeta_K\right)\,.
\end{align}

\subsection{Orbit of the extended body}

In order to describe the motion of the extended body according to the MPD
model, we need both the timelike unit vector $U$ tangent to the center world
line and the unit timelike vector $u$ aligned with the 4-momentum. In the
following, we shall assume that the world line of the extended body is
confined onto the equatorial plane, so that the 4-velocity $U$ is given by
Eq.~\eqref{polarnu}, with
\beq
\label{polarnu2}
\nu(U,n) \equiv \nu \hat{\nu} \equiv \nu^{\hat r}e_{\hat r}
+\nu^{\hat \phi}e_{\hat \phi} 
=  \nu (\cos \alpha e_{\hat r}+ \sin \alpha e_{\hat \phi})\,,
\eeq
where $\nu$ and $\alpha\in[0,\frac{\pi}{2}]$ are the signed magnitude of the
spatial velocity and its polar angle, measured clockwise from the positive
$\phi$ direction in the $r$-$\phi$ tangent plane, respectively, while
$\hat\nu\equiv\hat \nu(U,n)$ is the associated unit vector; hence, $\nu$ has
positive/negative values for co/counter-rotating azimuthal motion
($\alpha=\pi/2$) and outward/inward radial motion ($\alpha=0$) with respect to
the ZAMOs, respectively.

A similar decomposition holds for the (body) 4-momentum $P=mu$, in the case of
equatorial orbits:
\beq
\label{polarnuu}
u=\gamma_u [n +\nu_u\hat \nu_u]\,, \qquad
\gamma_u=(1-\nu_u^2)^{-1/2}\,,
\eeq
with
\beq
\label{polarnuu2}
\hat \nu (u,n)\equiv \hat \nu_u=\cos\alpha_u e_{\hat r}
+ \sin\alpha_u e_{\hat \phi}\,,
\eeq
and $\alpha_u\in[0,\frac{\pi}{2}]$. An orthonormal frame adapted to $u\equiv
e_0$ can then be built by introducing the spatial triad:
\beq
\label{uframe}
e_1\equiv \hat \nu_u^\perp \equiv \sin\alpha_u e_{\hat r}
- \cos\alpha_u e_{\hat \phi}\,,\quad
e_2={\rm sgn}(\nu_u)\gamma_u (\nu_u n +\hat \nu_u)\,,\quad
e_3=-e_{\hat \theta}\,.
\eeq
The dual frame of $\{e_\alpha\}$ will be referred to as $\{\omega^\alpha\}$, with
$\omega^0=-u^\flat$, $u^\flat$ being the covariant dual of $u$.
The projection of the spin vector into the local rest space of $u$ defines the
spin vector $S(u)$ (hereafter simply denoted by $S$, for short).
In the frame \eqref{uframe}, the spin $S$ decomposes as
\beq
S=S^1e_1+S^2e_2+S^3e_3\,.
\eeq

\subsection{Setting the body's spin and quadrupole in the aligned case}

In the following, we shall consider the special case where the spin vector is
aligned with the spacetime rotation axis, i.e.,
\beq
S=se_3\,.
\eeq
This entails that the spin and quadrupole terms entering the right-hand sides
of Eqs.~\eqref{papcoreqs1} and~\eqref{papcoreqs2} decompose, with respect to
the frame adapted to $u$, as
\begin{align}
\label{fspinframeu}
F_{\rm (spin)}&=F_{\rm (spin)}^0u+F_{\rm (spin)}^1e_1+F_{\rm (spin)}^2e_2\,,\nonumber\\
F_{\rm (quad)}&=F_{\rm (quad)}^0u+F_{\rm (quad)}^1e_1+F_{\rm (quad)}^2e_2\,,
\end{align}
and
\begin{align}
\label{dspinframeu}
D_{\rm (spin)}&=-\omega^0\wedge{\mathcal E}_{\rm (spin)}(u)\,,\nonumber\\
D_{\rm (quad)}&=-\omega^0\wedge{\mathcal E}_{\rm (quad)}(u)\,,
\end{align}
with
\begin{align}
\label{dspinframeu2}
{\mathcal E}_{\rm (spin)}(u)&={\mathcal E}_{\rm (spin)}{}_1\omega^{1}
+{\mathcal E}_{\rm (spin)}{}_2\omega^{2}\,,\nonumber\\
{\mathcal E}_{\rm (quad)}(u)&={\mathcal E}_{\rm (quad)}{}_1 \omega^1 
+{\mathcal E}_{\rm (quad)}{}_2 \omega^2\,,
\end{align}
respectively. The explicit expressions for the above components are listed in
Appendix~\ref{spinandquadframecompts}.


\section{Solving the MPD equations for non-precessing equatorial
  orbits}

\subsection{Complete set of evolution equations}

Under the assumptions of equatorial motion and aligned spins discussed in the
previous section the whole set of MPD equations
\eqref{papcoreqs1}--\eqref{tulczconds} reduces to
\begin{align}
\label{setfin}
\frac{\rmd m}{\rmd \tau} &= 
F_{\rm (spin)}^0 + F_{\rm (quad)}^0\,,\nonumber\\
\frac{\rmd \alpha_u}{\rmd \tau} &= 
-\frac{\gamma}{\nu_u}\left[\nu\cos(\alpha_u+\alpha)
-\nu_u\right]\theta_{\hat\phi}(n)^{\hat r}
+\frac{\gamma}{\nu_u}\left(\sin\alpha_ua(n)^{\hat r}
+\nu\nu_u\sin\alpha \, k_{\rm (Lie)}\right)\nonumber\\
&
-\frac{1}{m\gamma_u\nu_u}(F_{\rm (spin)}^1 + F_{\rm (quad)}^1)\,,\nonumber\\
\frac{\rmd \nu_u}{\rmd \tau} &=
-\frac{\gamma}{\gamma_u^2}\left[\cos\alpha_ua(n)^{\hat r}
+\nu\sin(\alpha_u+\alpha)\theta_{\hat\phi}(n)^{\hat r}\right]
+\frac{1}{m\gamma_u^2}(F_{\rm (spin)}^2 + F_{\rm (quad)}^2)\,,\nonumber\\
\frac{\rmd s}{\rmd \tau} &=0\,,
\end{align}
together with the compatibility conditions
\beq
\label{compatib}
0=e_3\times({\mathcal E}_{\rm (spin)}(u)+{\mathcal E}_{\rm (quad)}(u)) 
+\frac{s}{m}(F_{\rm (spin)} + F_{\rm (quad)})\,,
\eeq
which come from the spin evolution equations and yield two algebraic relations
for $\nu$ and $\alpha$. The last equation~\eqref{setfin} implies that the
signed spin magnitude $s$ is a constant of motion. Finally, Eqs.~\eqref{setfin}
must be coupled with the decomposition~\eqref{polarnu} and~\eqref{polarnu2} of
$U=\rmd x^\alpha/\rmd\tau$ to provide the remaining unknowns $t$, $r$ and
$\phi$ [see also Eqs.~\eqref{Uequatgen}]. Eqs.~\eqref{setfin}
and~\eqref{Uequatgen} are written in a form that is suitable for the numerical
integration. Additional (non-independent) relations are obtained from the
conservation~\eqref{totalenergy} of the total energy and angular momentum,
\begin{align}
\label{EandJ}
E&=N\gamma_u\left[m+s(\nu_u\sin\alpha_u a(n)^{\hat r}
+\theta_{\hat\phi}(n)^{\hat r})\right]-N^\phi J
\,,\nonumber\\
J&=\gamma_u\sqrt{g_{\phi\phi}}\left[m\nu_u\sin\alpha_u-s(k_{\rm (Lie)}
+\nu_u\sin\alpha_u\theta_{\hat\phi}(n)^{\hat r})\right]
\,.
\end{align}
These can be used as a consistency check. Examples of numerically-integrated
orbits are discussed in Refs. \cite{quadrupkerr1,quadrupkerrnum}.

We are interested here in studying the general features of equatorial motion
to second order in spin, taking advantage of the simplified form of the MPD
equations discussed in Section \ref{simplifiedMPD}. We shall also derive
analytic solutions for the orbits to that order by computing the corrections,
produced by a non-zero spin, to a reference circular geodesic motion.

\subsection{Solution to the order $\boldsymbol{\mathcal{O}(S^2)}$}

Let us introduce the dimensionless spin parameter
\beq
{\hat s}=\frac{s}{m_0M}\,,
\eeq
where $m_0$ denotes the ``bare'' mass of the body. We shall systematically
neglect terms that are of orders higher than the second in ${\hat s}$,
hereafter. Hence, all quantities must be understood as being evaluated up to
the order $\mathcal{O}(\hat{s}^2)$. Our set of equations can then be
simplified by means of Eqs.~\eqref{mJdef} and~\eqref{reluUgen}, which yield
\beq
\label{mdef}
m=m_0-\frac14C_Qm_0M^2{\hat s}^2\gamma_u^2\left[
(2E_{\hat r\hat r}+E_{\hat \theta\hat \theta})\nu_u^2\cos2\alpha_u
+(2+\nu_u^2)E_{\hat \theta\hat \theta}
+4\nu_u\sin\alpha_uH_{\hat r\hat \theta}
\right]\,, 
\eeq
and
\begin{align}
\label{compatib2}
\nu&=\nu_u+\frac12\left(1-C_Q\right)M^2{\hat s}^2\left\{
\nu_u\left[(2E_{\hat r\hat r}+E_{\hat \theta\hat \theta})\cos2\alpha_u
+3E_{\hat \theta\hat \theta}\right]
+2H_{\hat r\hat \theta}(1+\nu_u^2)\sin\alpha_u
\right\}\,,\nonumber\\
\alpha&=\alpha_u-
\left(1-C_Q\right)M^2{\hat s}^2\frac{\cos\alpha_u}{\nu_u}\left[
(2E_{\hat r\hat r}+E_{\hat \theta\hat \theta})\nu_u\sin\alpha_u
-H_{\hat r\hat \theta}\right]\,,
\end{align}
respectively. Here, $E_{\hat r\hat r}$, $E_{\hat \theta\hat \theta}$ are
components of the electric part of the Riemann tensor, while $H_{\hat r\hat
  \theta}$ is a component of its magnetic part (see Appendix~\ref{appzamos}).
Eqs.~\eqref{compatib2} show that the value of the polarizability
for black holes, namely $C_Q=1$, is a very special case leading to a great
simplification (see below). 

\begin{figure} 
\typeout{*** EPS/PDF figure 1}
\begin{center}
\includegraphics[scale=0.3]{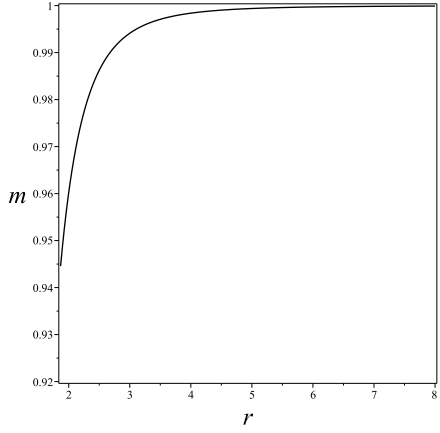}
\end{center}
\caption{The behavior of the mass $m$ of the body (in units of $m_0$) as a
  function of the radial coordinate (in units of $M$) is shown for the
  following choice of parameters: $a/M=0.5$, $C_Q=1$, ${\hat s}=0.25$,
  $E/m_0=1$ and $J/(m_0M)=4$. The mass shift in this case equates
  $|m(+\infty)-m(r_+)|\approx0.055m_0$. 
	The value of the dimensionless spin parameter $\hat s$ has been
    exaggerated in order to enhance the effect.}
\label{fig:mdir}
\end{figure}

On the other hand, solving Eqs.~\eqref{EandJ} algebraically for $\nu_u$ and
$\alpha_u$ leads to
\begin{align}
\label{nuusol}
\qquad\gamma_u&=\frac{E+N^\phi J}{m_0N}\left\{
1-M{\hat s}\left[\theta_{\hat\phi}(n)^{\hat r}
+\frac{Na(n)^{\hat r}J}{(E+N^\phi J)\sqrt{g_{\phi\phi}}}\right]
-M^2{\hat s}^2\left(\frac{m_2}{m_0M^2}+E_{\hat r\hat r}
+E_{\hat \theta\hat \theta}\right)\right\}\,,\nonumber\\
\gamma_u\nu_u\sin\alpha_u&=
\frac{J}{m_0\sqrt{g_{\phi\phi}}}\left\{
1+M{\hat s}\left[\theta_{\hat\phi}(n)^{\hat r}
+(E+N^\phi J)\frac{\sqrt{g_{\phi\phi}}k_{\rm (Lie)}}{NJ}\right]
-M^2{\hat s}^2\left(\frac{m_2}{m_0M^2}+E_{\hat r\hat r}
+E_{\hat \theta\hat \theta}\right)
\right\}\,.
\end{align}
Those identities can be used next at the lowest order in the previous
equations so as to express $m$, $\nu$ and $\alpha$ in terms of $E$ and $J$.
In particular, the solution for the body mass is found to be
\beq
\label{solmr}
m = m_0\left\{1-\frac12C_Q\frac{M^3}{r^3}
\left[1+\frac{3}{m_0^2r^2}(J-aE)^2\right]{\hat s}^2\right\}
\equiv m_0+{\hat s}^2m_2\,.
\eeq
Its behavior as a function of the radial coordinate is shown in
Fig.~\ref{fig:mdir} for selected values of the parameters. It is interesting
to evaluate the difference between the limiting values at the horizon and at
infinity, $|m(+\infty)-m(r_+)|={\hat s}^2m_2(r_+)$, since this represents the
largest mass variation during the evolution.

It can be verified that the solution~\eqref{compatib2} for $U$ is
equivalently obtained from the compatibility conditions~\eqref{compatib}.
Furthermore, the truncated equation~\eqref{setfin} for the mass reads
\beq
\label{eqmass}
\frac{\rmd m}{\rmd \tau} = F_{\rm (quad)}^0+\mathcal{O}({\hat s}^3)\,,
\eeq
with
\begin{align}
\label{Fquad0}
F_{\rm (quad)}^0=
\frac{1}{2}m_0M^2{\hat s}^2C_Q\gamma_u\nu_u\cos\alpha_u
\left\{\gamma_u^2\nu_u\left[
\nu_u(b_1\cos2\alpha_u+c_3)+2(a_1-b_4)\sin\alpha_u\right]
+\frac13(2c_1-c_2)
\right\}
\,,
\end{align}
where the lowest order piece of the solution~\eqref{nuusol} may be used, which
implies notably that $F_{\rm (spin)}^0=\mathcal{O}({\hat s}^3)$. The
quantities $a_i$, $b_i$ and $c_i$, all functions of $r$, are listed in
Appendix~\ref{spinandquadframecompts}. Once parametrized with the radial
coordinate $r$ instead of the proper time $\tau$ by means of
Eqs.~\eqref{Uequatgen}, formula~\eqref{eqmass} takes the very simple form
\beq
\label{eqmassr}
\frac{\rmd m_2}{\rmd r} = \frac{3}{2}m_0C_Q\frac{M^3}{r^4}
\left[1+\frac{5}{m_0^2r^2}(J-aE)^2\right]\,,
\eeq
whose solution coincides with that of Eq.~\eqref{solmr}.


%

\subsection{Circular motion}

\subsubsection{Solution to the order $\mathcal{O}({\hat s}^2)$}

In this subsection, we restrict ourselves to circular orbits, as described by
the parametrizations~\eqref{orbita}--\eqref{casocircorb}. For circular motion
in the equatorial plane, we must set $\alpha=\pi/2$, so that
Eqs.~\eqref{compatib2} become
\begin{align}
\label{compatibcirc}
\nu&=\nu_u+\left(1-C_Q\right)M^2{\hat s}^2\left[
-\nu_\pm(E_{\hat r\hat r}-E_{\hat \theta\hat \theta})+H_{\hat r\hat \theta}(1+\nu_\pm^2)
\right]\,,\nonumber\\
\alpha_u&=\frac{\pi}{2}\,,
\end{align}
to second order in $\hat s$. Thus, $F_{\rm (quad)}^0=0$ from its
expression~\eqref{Fquad0}, and Eq.~\eqref{eqmass} tells us that the mass of
the body is a constant of motion to that order. However, it differs from the
bare mass $m_0$ by virtue of the general definition~\eqref{mdef}, which yields
\begin{align}
\label{mcirc}
m&=m_0-\frac12C_Qm_0M^2{\hat s}^2\gamma_\pm^2\left[
E_{\hat \theta\hat \theta}-\nu_\pm^2E_{\hat r\hat r}
+2\nu_\pm H_{\hat r\hat \theta}
\right]\nonumber\\
&=m_0\left[1-\frac12C_Q\frac{M^3}{r^{7/2}}
\frac{r^2\mp4a\sqrt{Mr}+3a^2}{r^{3/2}
-3M\sqrt{r}\pm2a\sqrt{M}}{\hat s}^2\right]\,,
\end{align}
once evaluated at $\alpha_u=\pi/2$, with $\nu_u$ replaced by $\nu_\pm$ to
the lowest order.

\begin{figure*} 
\typeout{*** EPS/PDF figure 5}
\begin{center}
$\begin{array}{cc}
\includegraphics[scale=0.3]{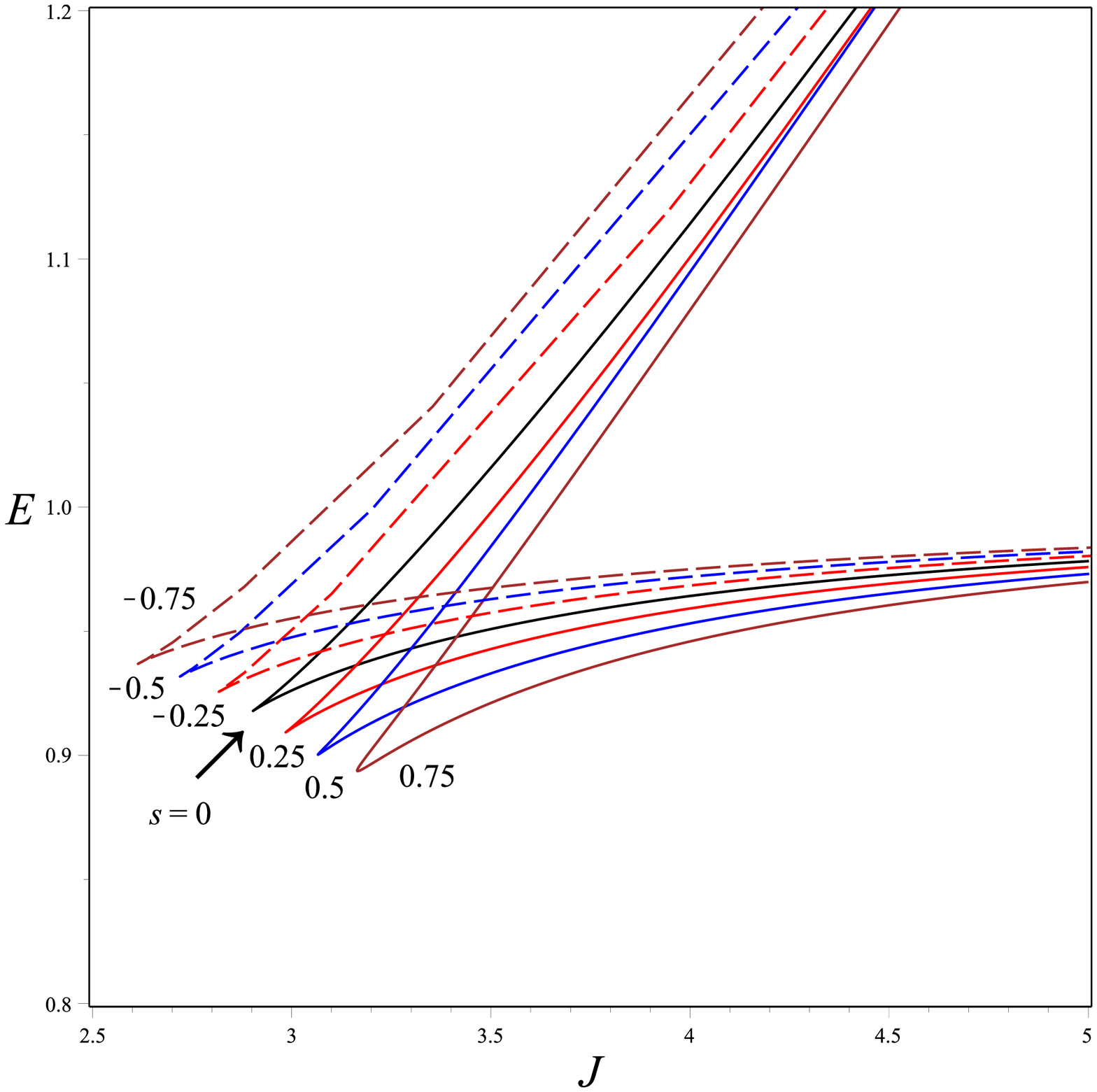}&\quad
\includegraphics[scale=0.3]{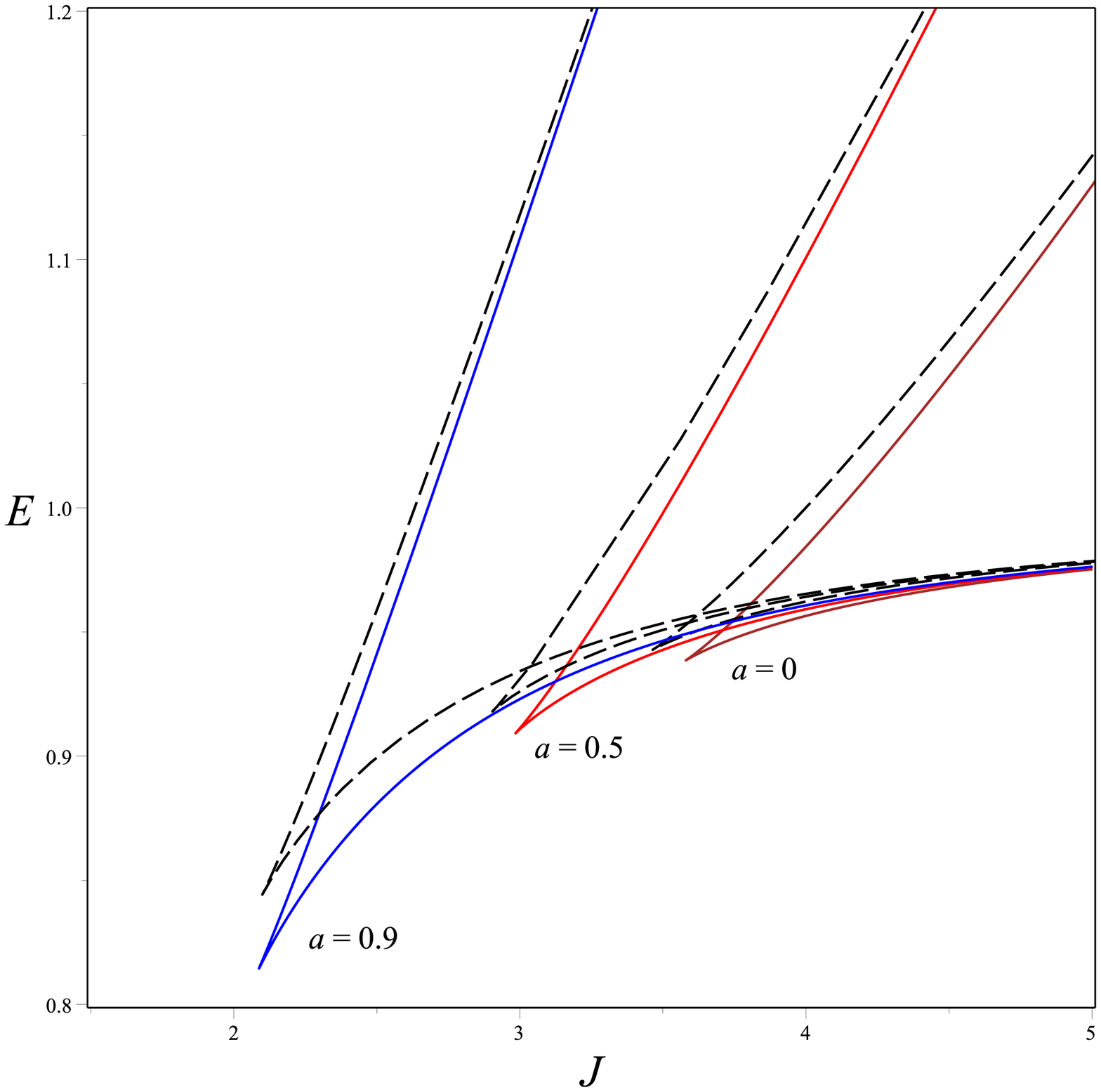}\\[.4cm]
\quad\mbox{(a)}\quad &\quad \mbox{(b)}
\end{array}$\\
\end{center}
\caption{The behavior of the energy $E$ versus the angular momentum $J$ in the
  case of co-rotating circular orbits is shown in panel (a), for the fixed
  black-hole dimensionless spin parameter $a/M=0.5$ and different values of
  $\hat s$; in panel (b), for the fixed body spin $\hat s=0.25$ and different
  values of $a/M$; $C_Q=1$ in both cases. Dashed curves in (b) correspond to
  the geodesic motion ($\hat s=0$).}
\label{fig:E_vs_J}
\end{figure*}

Next, solving algebraically the second equation~\eqref{setfin} for $\nu_u$ we
find
\beq
\nu_u=\nu_\pm+{\hat s}\nu_{u(1)}+{\hat s}^2\nu_{u(2)}\,,
\eeq
where $\nu_\pm$ denote the geodesic linear velocities~\eqref{geocirc}, while
$\nu_{u(1)}$ and $\nu_{u(2)}$ are the spin-induced corrections
\begin{align}
\nu_{u(1)}&=\pm\frac{M}{2\zeta_K}[-\nu_\pm(E_{\hat r\hat r}-E_{\hat \theta\hat \theta})
+(1+\nu_\pm^2)H_{\hat r\hat \theta}]\,,\nonumber\\
\nu_{u(2)}&=\mp\frac{M^2}{\gamma_\pm^2\zeta_K}{\tilde F}^1_{\rm (quad)}\nonumber\\
&
+\nu_{u(1)}\left[\nu_{u(1)}\left(\pm\frac{k_{\rm (Lie)}}{2\zeta_K}
+\frac1{\nu_\pm}\right)
\pm\left(1-C_Q\right)M\zeta_K\pm
\frac{MH_{\hat r\hat \theta}}{2\gamma_\pm^2\nu_\pm\zeta_K}\right]\,,
\end{align}
of first and second orders, respectively, with (see also Eq.~\eqref{eqE5})
\beq
{\tilde F}_{\rm (quad)}^1=\frac{1}{4}C_Q\left\{
\gamma_\pm^2[b_1-b_2+\nu_\pm(b_4-b_5)]-b_1+\frac13(b_2+b_3)
\right\}\,.
\eeq
Substituting the above solutions for $\nu_u$ and $\alpha_u$ into
Eq.~\eqref{EandJ} we obtain
\begin{align}
\label{EandJcirc}
E&=m_0N\gamma_\pm\left(1-\frac{\sqrt{g_{\phi\phi}}N^\phi}{N}\nu_\pm\right)
\nonumber\\ &
+m_0\gamma_\pm\left[
\gamma_\pm^2\zeta_\pm\nu_{u(1)}\sqrt{g_{\phi\phi}}
+MN(\nu_\pm a(n)^{\hat r}+\theta_{\hat\phi}(n)^{\hat r})
+M\sqrt{g_{\phi\phi}}N^\phi(\nu_\pm\theta_{\hat\phi}(n)^{\hat r}+k_{\rm (Lie)})
\right]{\hat s}\nonumber\\
&
+\left\{
m_2N\gamma_\pm\left(1-\frac{\sqrt{g_{\phi\phi}}N^\phi}{N}\nu_\pm\right)
+m_0\gamma_\pm^3\left[
\sqrt{g_{\phi\phi}}\zeta_\pm\left(\pm M\zeta_K\nu_{u(1)}+\nu_{u(2)}
+\frac32\gamma_\pm^2\nu_\pm\nu_{u(1)}^2\right)
+\frac12N\nu_{u(1)}^2
\right]
\right\}{\hat s}^2
\,,\nonumber\\
J&=m_0\gamma_\pm\nu_\pm\sqrt{g_{\phi\phi}}\nonumber\\
&
+m_0\gamma_\pm\sqrt{g_{\phi\phi}}\left[\gamma_\pm^2\nu_{u(1)}
-M(\nu_\pm\theta_{\hat\phi}(n)^{\hat r}+k_{\rm (Lie)})
\right]{\hat s}\nonumber\\
&
+\left[
m_2\gamma_\pm\nu_\pm\sqrt{g_{\phi\phi}}
+m_0\gamma_\pm^3\sqrt{g_{\phi\phi}}\left(\pm M\zeta_K\nu_{u(1)}+\nu_{u(2)}
+\frac32\gamma_\pm^2\nu_\pm\nu_{u(1)}^2\right)
\right]{\hat s}^2
\,.
\end{align}
The behavior of the energy $E$ versus the angular momentum $J$ in the case of
co-rotating orbits is shown in Fig. \ref{fig:E_vs_J} for selected values of
the parameters.

Finally, the 4-velocity $U$ is given by Eq.~\eqref{orbita}, with normalization
factor
\beq
\label{Gammaquad}
\Gamma=\Gamma_\pm\left\{
1+\gamma_\pm^2\nu_\pm\nu_1{\hat s}+\gamma_\pm^2\left[\nu_\pm\nu_2
+\left(\frac32\gamma_\pm^2-1\right)\nu_1^2\right]{\hat s}^2
\right\}\,,
\eeq
and angular velocity  
\beq
\label{zetaquad}
\zeta=\zeta_\pm\left[1+\frac{N}{\zeta_\pm\sqrt{g_{\phi\phi}}}(\nu_1{\hat s}
+\nu_2{\hat s}^2)\right]\,,
\eeq
where
\beq
\nu_1=\nu_{u(1)}\,,\qquad
\nu_2=\nu_{u(2)}\pm2\left(1-C_Q\right)M\zeta_K\nu_1\,.
\eeq
The relation between the timelike unit tangent vector $U$ to the body center
world line and the unit timelike vector $u$ aligned with the 4-momentum thus
reads
\beq
\label{relUu}
U-u={\hat s}^2\gamma_\pm^2\left(1-C_Q\right)M\zeta_K\nu_1\bar U_{\pm}\,,
\eeq
the unit vector $\bar U_{\pm}$ being already defined in Eq.~\eqref{barUgeo}.
Hence, in general, $U$ and $P$ are not aligned unless $C_Q=1$, as discussed
below.

\subsubsection{Weak field limit}

Let us study now the weak field limit of the above analysis. For convenience,
we introduce the dimensionless quantities $u_0=M/r_0$ and $\hat a=a/M$.
We may consider only the case of co-rotating orbits, the counter-rotating
case simply following from the replacement $\hat a\to-\hat a$. Every quantity
is expanded up to a certain power of $u_0$ as follows
\beq
X\sim X_0+{\hat a}X_{\hat a}+{\hat s}X_{\hat s}
+{\hat a}^2X_{\hat a\hat a}+2{\hat a}{\hat s}X_{\hat a\hat s}
+{\hat s}^2X_{\hat s\hat s}\,,
\eeq
where terms of orders higher than the second in the background rotation
parameter, as well as terms like $\hat a\hat s^2$ and $\hat a^2\hat s$, are
neglected.

The weak field expansion of the conserved energy and angular
momentum~\eqref{EandJcirc} are then found to be
\begin{align}
\frac{E}{m_0}&=1-\frac12u_0+\frac{3}{8}u_0^2+\frac{27}{16}u_0^3
+\frac{675}{128}u_0^4
-\frac{1}{16}u_0^{5/2}(8+36u_0+135u_0^2)(2{\hat a}+{\hat s}) \nonumber\\
&
+\frac14u_0^3(2+15u_0){\hat a}^2
+\frac14u_0^3(2+23u_0){\hat a}{\hat s}
+\frac18u_0^3[2C_Q-(6-17C_Q)u_0]{\hat s}^2 
+ \mathcal{O}(u_0^5) 
\,,\nonumber\\
\frac{J}{m_0M}&=u_0^{-1/2}\left(1+\frac32u_0+\frac{27}{8}u_0^2
+\frac{135}{16}u_0^3+\frac{2835}{128}u_0^4+\frac{15309}{256}u_0^5\right)
-3u_0\left(1+\frac52u_0+\frac{63}{8}u_0^2+\frac{405}{16}u_0^3\right)
{\hat a} \nonumber\\
&
+\left(1-2u_0-\frac{27}{8}u_0^2-\frac{81}{8}u_0^3
-\frac{4185}{128}u_0^4\right){\hat s}
+u_0^{3/2}\left(1+5u_0+\frac{189}{8}u_0^2
+\frac{405}{4}u_0^3\right){\hat a}^2\nonumber\\
&
+\frac32u_0^{3/2}\left(1+\frac92u_0+\frac{159}{8}u_0^2
+\frac{1305}{16}u_0^3\right){\hat a}{\hat s}\nonumber\\
&
+\frac34u_0^{3/2}\left[C_Q-\frac12\left(5-\frac{23}{3}C_Q\right)u_0
-\frac14\left(7-\frac{79}{2}C_Q\right)u_0^2
+\frac{1}{16}(81+477C_Q)u_0^3\right]{\hat s}^2
+ \mathcal{O}(u_0^5) \,,
\end{align}
whereas the normalization factor~\eqref{Gammaquad} and the angular
velocity~\eqref{zetaquad} become
\begin{align}
\Gamma&=1+\frac32u_0+\frac{27}{8}u_0^2+\frac{135}{16}u_0^3
+\frac{2835}{128}u_0^4
-\frac{3}{16}u_0^{5/2}(8+36u_0+135u_0^2)(2{\hat a}+{\hat s})\nonumber\\
&
+\frac14u_0^3(2+27u_0){\hat a}^2
+\frac34u_0^3(2+19u_0){\hat a}{\hat s}
+\frac38u_0^3[2C_Q+5(2+C_Q)u_0]{\hat s}^2 
+ \mathcal{O}(u_0^5)
\,,\nonumber\\
M\zeta&=u_0^{3/2}-\left({\hat a}+\frac32{\hat s}\right)u_0^3
+u_0^{9/2}{\hat a}^2
+\frac32u_0^{7/2}(1+2u_0){\hat a}{\hat s}
+\frac34u_0^{7/2}\left[C_Q+\left(\frac{7}{2}-2C_Q\right)\right]{\hat s}^2
+ \mathcal{O}(u_0^5)\,,
\end{align}
respectively. In order to derive gauge invariant expressions, we express
$E$ and $J$ in terms of the gauge-invariant dimensionless variable
$y=(M\zeta)^{2/3}$, related to $u_0$ by
\beq
u_0=y+\left(\frac23{\hat a}+{\hat s}\right)y^{5/2}
+\frac59y^4{\hat a}^2
-y^3\left(1-\frac53y\right){\hat a}{\hat s}
-\frac12y^3[C_Q-2(1+C_Q)y]{\hat s}^2 
+ \mathcal{O}(y^{9/2}) \,.
\eeq 
This yields
\begin{align}
\frac{E}{m_0}&=1-\frac12y+\frac38y^2+\frac{27}{16}y^3+\frac{675}{128}y^4
-y^{5/2}\left(\frac43+4y+\frac{27}{2}y^2\right){\hat a}
-y^{5/2}\left(1+\frac32y+\frac{27}{8}y^2\right){\hat s}\nonumber\\
&
+\frac12y^3\left(1+\frac{65}{18}y\right){\hat a}^2
+y^3\left(1+\frac56y\right){\hat a}{\hat s}
+\frac12y^3\left[C_Q
-5\left(1-\frac12C_Q\right)y
\right]{\hat s}^2 
+ \mathcal{O}(y^5)
\,,\nonumber\\
\frac{J}{m_0M}&=y^{-1/2}\left(1+\frac32y+\frac{27}{8}y^2+\frac{135}{16}y^3
+\frac{2835}{128}y^4+\frac{15309}{256}y^5\right)
-y\left(\frac{10}{3}+7y+\frac{81}{4}y^2-\frac{495}{8}y^3\right){\hat a}
\nonumber\\
&
+\left(1-\frac52y-\frac{21}{8}y^2-\frac{81}{16}y^3
-\frac{1485}{128}y^4\right){\hat s}
+y^{3/2}\left(1+\frac{26}{9}y+\frac{335}{24}y^2
+\frac{459}{8}y^3\right){\hat a}^2
\nonumber\\
&
+y^{3/2}\left(2+\frac43y+\frac{25}{4}y^2
+\frac{81}{4}y^3\right){\hat a}{\hat s}\nonumber\\
&
+y^{3/2}\left[C_Q
-4\left(1-\frac{1}{2}C_Q\right)y
-\frac{15}{2}\left(1-\frac{3}{4}C_Q\right)y^2
-\frac{81}{4}\left(1-\frac{5}{6}C_Q\right)y^3
\right]{\hat s}^2 
+ \mathcal{O}(y^5) \,.
\end{align}

\subsubsection{Comparison with Ref.~\cite{hinderer}}

Before investigating the general equatorial motion, let us show how the above
analysis allows us to reproduce the results of Ref.~\cite{hinderer}. As
already stated, our quadrupole tensor reduces to the one adopted there to
describe black holes if we set $C_Q=1$. Since we have then
$\nu=\nu_u+\mathcal{O}({\hat s}^3)$ from Eq.~\eqref{compatibcirc} [see also
Eq.~\eqref{relUu}], this entails $P\propto U+\mathcal{O}({\hat s}^3)$. In
order to write our expressions for the energy and angular momentum in the same
form as in Ref.~\cite{hinderer}, we eliminate the dependence on the radial
coordinate in favor of the angular velocity by inverting perturbatively
Eq.~\eqref{zetaquad}:
\begin{align}
r=r_c
+\frac{M}{r_c}(-\sqrt{Mr_c}\pm a){\hat s}
+\frac{M^2}{2r_c^3}\bigg\{
\Delta_c-2a(a\mp\sqrt{Mr_c})-\left(1-C_Q\right)[\Delta_c+4a(a\mp\sqrt{Mr_c})]
\bigg\}{\hat s}^2\,,
\end{align}
where $r_c^{3/2}=\sqrt{M}\left(\frac1\zeta\mp a\right)$ and 
$\Delta_c = \Delta(r=r_c)$.
Substituting next into Eq.~\eqref{EandJcirc} leads to
\begin{align}
\label{EandJcirc_rc}
E&=\frac{m_0}{r_c^{3/4}}
\frac{r_c^{3/2}-2M\sqrt{r_c}\pm a\sqrt{M}}{\left[r_c^{3/2}-3M\sqrt{r_c}
\pm2a\sqrt{M}\right]^{1/2}}\mp\frac{m_0M^2}{r_c^{9/4}}
\frac{\sqrt{Mr_c}\mp a}{\left[r_c^{3/2}
-3M\sqrt{r_c}\pm2a\sqrt{M}\right]^{1/2}}{\hat s}
\nonumber\\
&
+\frac{m_0M^3}{2r_c^{15/4}}
\frac{1}{\left[r_c^{3/2}-3M\sqrt{r_c}\pm2a\sqrt{M}\right]^{1/2}}\Bigg\{
r_c^{3/2}-4M\sqrt{r_c}\pm3a\sqrt{M}\nonumber\\
&
\qquad \qquad -\left(1-C_Q\right)\left[
r_c^{3/2}+M\sqrt{r_c}\mp9a\sqrt{M}+\frac{11a^2}{2\sqrt{r_c}}
+\frac32
\frac{(r_c+3M)a^2+2M^{3/2}\sqrt{r_c}(\sqrt{Mr_c}\mp3a)}{r_c^{3/2}
-3M\sqrt{r_c}\pm2a\sqrt{M}}
\right] \Bigg\}{\hat s}^2\,,\nonumber\\
J&=\pm\frac{m_0\sqrt{M}}{r_c^{3/4}}
\frac{r_c^2\mp2a\sqrt{Mr_c}+a^2}{\left[r_c^{3/2}
-3M\sqrt{r_c}\pm2a\sqrt{M}\right]^{1/2}}
+\frac{m_0M}{r_c^{9/4}}
\frac{r_c^2(r_c-4M)+Ma^2\pm a\sqrt{Mr_c}(3r_c-M)}{\left[r_c^{3/2}
-3M\sqrt{r_c}\pm2a\sqrt{M}\right]^{1/2}}{\hat s}\nonumber\\
&
\pm\frac{m_0M^{5/2}}{2r_c^{15/4}}\frac{1}{\left[r_c^{3/2}
-3M\sqrt{r_c}\pm2a\sqrt{M}\right]^{1/2}}\Bigg\{
3Ma^2\pm2a\sqrt{Mr_c}(3r_c-2M)+r_c^2(2r_c-7M)\nonumber\\
&\left. \qquad \qquad
-\left(1-C_Q\right)\left[
(2r_c+3M)(r_c^2-Mr_c+3M^2)\mp4a\sqrt{Mr_c}(3r_c+2M)
+10r_ca^2\pm\frac{11\sqrt{M}a^3}{2\sqrt{r_c}}
\right.\right.\nonumber\\
&\left. \qquad \qquad 
+\frac32\sqrt{M}\frac{6M^{3/2}\sqrt{r_c}(3M^2+4a^2)
\mp a[a^2(r_c+9M)+4M^2(5r_c+3M)]}{r_c^{3/2}-3M\sqrt{r_c}\pm2a\sqrt{M}}
\right]
\Bigg\}{\hat s}^2\,.
\end{align}
This exactly reproduces the results of Ref.~\cite{hinderer} when specialized
to the case $C_Q=1$ [see their counterparts displayed in Eqs.~(39a) and~(39b)
there, with in addition $M=1=m_0$].


%

\subsection{General equatorial motion}

\subsubsection{Effective potential}

\begin{figure} 
\typeout{*** EPS/PDF figure 3}
\begin{center}
\includegraphics[scale=0.3]{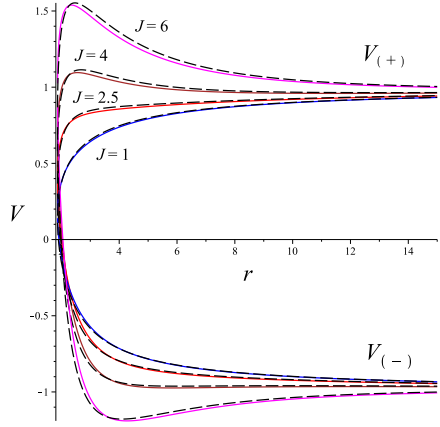}
\end{center}
\caption{The behavior of the effective potential for radial motion is shown
  for the following choice of parameters: $a/M=0.5$, $C_Q=1$, $\hat s=0.25$,
  and for different values of the body dimensionless angular momentum,
  $J/(m_0M)=[1,2.5,4,6]$. The corresponding geodesic case ($\hat s=0$) is also
  shown for comparison (dashed curves).}
\label{fig:poteff}
\end{figure}

In order to discuss the general features of equatorial motion it is most
useful to introduce appropriate effective potentials \cite{LL,MTW}. The latter naturally
arise when factoring the expression of $({\rmd r}/{\rmd \tau})^2$ as a
polynomial in the energy $E$ of the test object. This factorisation takes the
form
\begin{equation} \label{eq:factorisation}
\left(\frac{\rmd r}{\rmd \tau}\right)^2 =\frac{\gamma^2\nu^2\cos^2\alpha}{g_{rr}}
={\mathcal A}\,{\mathcal P}(E)(E-V_{(+)})(E-V_{(-)})+O({\hat s}^3)\,,
\end{equation}
where the solutions for $\nu$ and $\alpha$ in terms of the conserved energy
and angular momentum are given by Eqs.~\eqref{compatib2}--\eqref{nuusol}. More
precisely, we find
\begin{align}
\label{eq:factorisation2}
{\mathcal A}&=\frac{1}{m_0^2 N^2 g_{rr}}
\bigg\{
1-2M{\hat s}\theta_{\hat\phi}(n)^{\hat r}
-M^2{\hat s}^2\bigg[
k_{\rm (Lie)}(k_{\rm (Lie)}-a(n)^{\hat r}) \nonumber \\ 
& \qquad \qquad \qquad \qquad 
+(2-C_Q)(2E_{\hat r\hat r}+E_{\hat \theta\hat \theta})
\frac{J^2}{m_0^2g_{\phi\phi}}
+(3-C_Q)E_{\hat \theta\hat \theta}+3E_{\hat r\hat r}
\bigg] \bigg\}
\,,\nonumber\\
{\mathcal P}(E)&=
1+\frac{M^2{\hat s}^2}{m_0^2N^2}(2-C_Q)(E+N^\phi J)\left[
(E_{\hat r\hat r}+2E_{\hat \theta\hat \theta})(E+N^\phi J)
+2\frac{N}{\sqrt{g_{\phi\phi}}}H_{\hat r\hat \theta}J
\right]
\,,
\end{align}
and
\begin{equation}
V_{(\pm)}=V_0^{(\pm)}+{\hat s}V_1^{(\pm)}+{\hat s}^2V_2^{(\pm)}\,,
\end{equation}
with
\begin{align}
\label{poteffsol}
V_0^{(\pm)}&=
-N^{\phi}J\pm\frac{N}{\sqrt{g_{\phi\phi}}}\sqrt{m_0^2g_{\phi\phi}+J^2}\,,
\nonumber\\
V_1^{(\pm)}&=
\frac{MN}{\sqrt{g_{\phi\phi}}}\left[
(a(n)^{\hat r}+k_{\rm (Lie)})J\pm\theta_{\hat\phi}(n)^{\hat r}
\frac{m_0^2g_{\phi\phi}+2J^2}{\sqrt{m_0^2g_{\phi\phi}+J^2}}
\right]\,,\nonumber\\
V_2^{(\pm)}&=
\pm\frac{M^2N}{2\sqrt{g_{\phi\phi}}}\bigg\{
\pm2[\theta_{\hat\phi}(n)^{\hat r}(a(n)^{\hat r}+3k_{\rm (Lie)})
-C_QH_{\hat r\hat \theta}]J\nonumber\\
& \qquad \qquad \quad
+\sqrt{m_0^2g_{\phi\phi}+J^2}\left[2k_{\rm (Lie)}(k_{\rm (Lie)}+3a(n)^{\hat r})
-(4-C_Q)E_{\hat r\hat r}-(4+C_Q)E_{\hat \theta\hat \theta}\right]
\bigg\}\nonumber\\
&
\mp\frac{M^2m_0^2N\sqrt{g_{\phi\phi}}}{2\sqrt{m_0^2g_{\phi\phi}+J^2}}\bigg\{
[k_{\rm (Lie)}(k_{\rm (Lie)}+3a(n)^{\hat r})
-(3-C_Q)E_{\hat r\hat r}-3E_{\hat \theta\hat \theta}]\nonumber\\
&
+\frac{m_0^2g_{\phi\phi}}{m_0^2g_{\phi\phi}+J^2}[k_{\rm (Lie)}a(n)^{\hat r}
-(E_{\hat r\hat r}+E_{\hat \theta\hat \theta})]
\bigg\}\,.
\end{align}
If $E$ and $J$ are kept fixed as $\hat{s}$ goes to zero, the leading order
approximation of $\mathcal{P}(E)$ is obtained by setting exactly $\hat{s}=0$
in Eq.~\eqref{eq:factorisation2}, which shows that $\mathcal{P}(E)$ is
necessarily positive in the domain of validity of the small spin expansion.
The solutions $E=V_{(\pm)}$ are generalizations of the radial effective
potentials for a test particle in a Kerr spacetime to the case of an extended
body with spin-induced quadrupole moment. Their behavior as a function of the
radial distance is shown in Fig.~\ref{fig:poteff} for selected values of the
parameters. The upper/lower branch corresponds to the $+$/$-$ sign in
Eq.~\eqref{poteffsol}.

On the other hand, since the equation $({\rmd r}/{\rmd \tau})^2=0$ is quartic
in $E$, in order to give a complete account of effective potentials we should
also consider the solutions to the equation ${\mathcal P}(E)=0$, or
equivalently,
\beq
\label{eq:PE}
1+\frac{3aM^3{\hat s}^2}{m_0^2r^5}(2-C_Q)(E+N^\phi J) \left[
a(E+N^\phi J)
-2\frac{N^2}{\Delta}(r^2+a^2)J
\right]=0\,.
\eeq
If $C_Q<2$, no real solutions for the energy exist. If $C_Q=2$, then
${\mathcal P}(E)=1$, irrespective of $E$ and $J$. If $C_Q>2$ (so excluding the
black hole case $C_Q=1$), the above equation admits two real solutions
$E=W_{(\pm)}$, perturbatively in ${\hat s}$, only if $E$ scales as
$1/\hat{s}$, with
\begin{equation}
W_{(\pm)}=\frac{m_0}{\hat a}\left[\pm\frac{1}{{\hat s}\sqrt{3(C_Q-2)}}
\left(\frac{r}{M}\right)^{5/2}
+  \frac{J}{m_0M}\right] +O({\hat s})\,.
\end{equation}
Note that the solutions $W_{(\pm)}$ diverge in both limits ${\hat a}\to0$ and
${\hat s}\to0$ for fixed values of the radial distance. Furthermore, for fixed
values of the dimensionless spin parameters, they indefinitely grow for large
$r$, exhibiting a monotonic behavior $W_{(\pm)}\sim\pm({r}/{M})^{5/2}$, so
that there cannot exist circular orbits associated with them. In the
following, we shall actually exclude the configurations for which
$E=\mathcal{O}(1/\hat{s})$ and ignore both potentials $W_{(\pm)}$.

\subsubsection{Circular orbits and ISCO}

Circular orbits correspond to the extremal points of $V_{(+)}$ and solving for
$J$ the resulting equation, $V_{(+)}'=0$, provides the associated angular
momentum. Now, as pointed out by Le Tiec et al.~\cite{TBB}, the shift of the
ISCO frequency due to the peturbation induced to the spacetime background
metric by the particle itself is an important strong-field benchmark.
Gravitational self-force theory has provided very accurate analytic
predictions for it in the case of a spinless body in motion along a circular
geodesic on a Schwarzschild background, at the first order in the symmetric
mass ratio of the two objects. In the present situation, because of its
spinning and quadrupolar structure, the body deviates from geodesic motion in
the way described by the MPD model. As a result, the last stable circular
orbit undergoes a shift in the frequency, made of terms proportional to the
spin as well as the quadrupole. Measuring this effect can therefore provide
relevant information on the structure of the body. Conversely, having
information about the spin and quadrupolar structure of the body allows one to
make predictions on the frequency of circular motion and its deviation from
the corresponding geodesic value.

Stability requires that $V_{(+)}''\le0$. For a spinless object the latter
condition boils down to
\beq
r^2-6Mr-3a^2\pm8a\sqrt{Mr}\geq0\,.
\eeq
The equality can be analytically solved for the radius $r_{\rm ms}$ of the
marginally stable orbit (or ISCO)~\cite{bardeen}:
\beq
r_{\rm ms}^{\rm Kerr}=M\left[3+Z_2\mp\sqrt{(3-Z_1)(3+Z_1+2Z_2)}\right]\,,
\eeq
where the upper/lower sign refers to co/counter-rotating orbits, with
\beq
Z_1=1+\left(1-{\hat a}^2\right)^{1/3}\left[\left(1-{\hat a}\right)^{1/3}
+\left(1+{\hat a}\right)^{1/3}\right]\,,\quad
Z_2=\sqrt{3{\hat a}^2+Z_1^2}\,.
\eeq
The latter quantities are even functions of ${\hat a}$ satisfying 
$Z_2\geq Z_1$, $Z_1\leq3$ and, for ${\hat a}=0$, $Z_1=3=Z_2$. For small values
of ${\hat a}$, the Kerr ISCO radius may be expanded as
\beq
\frac{r_{\rm ms}^{\rm Kerr}}{M}=6\mp\frac{4\sqrt{6}}{3}{\hat a}
-\frac{7}{18}{\hat a}^2\mp\frac{13\sqrt{6}}{162}{\hat a}^3
-\frac{241}{1944}{\hat a}^4+O({\hat a}^5)\,.
\eeq
We recover the value $r_{\rm ms}^{\rm Kerr}=6M$ in the Schwarzschild case
(${\hat a}=0$). 

\begin{figure} 
\typeout{*** EPS/PDF figure 4}
\begin{center}
$\begin{array}{cc}
\includegraphics[scale=0.3]{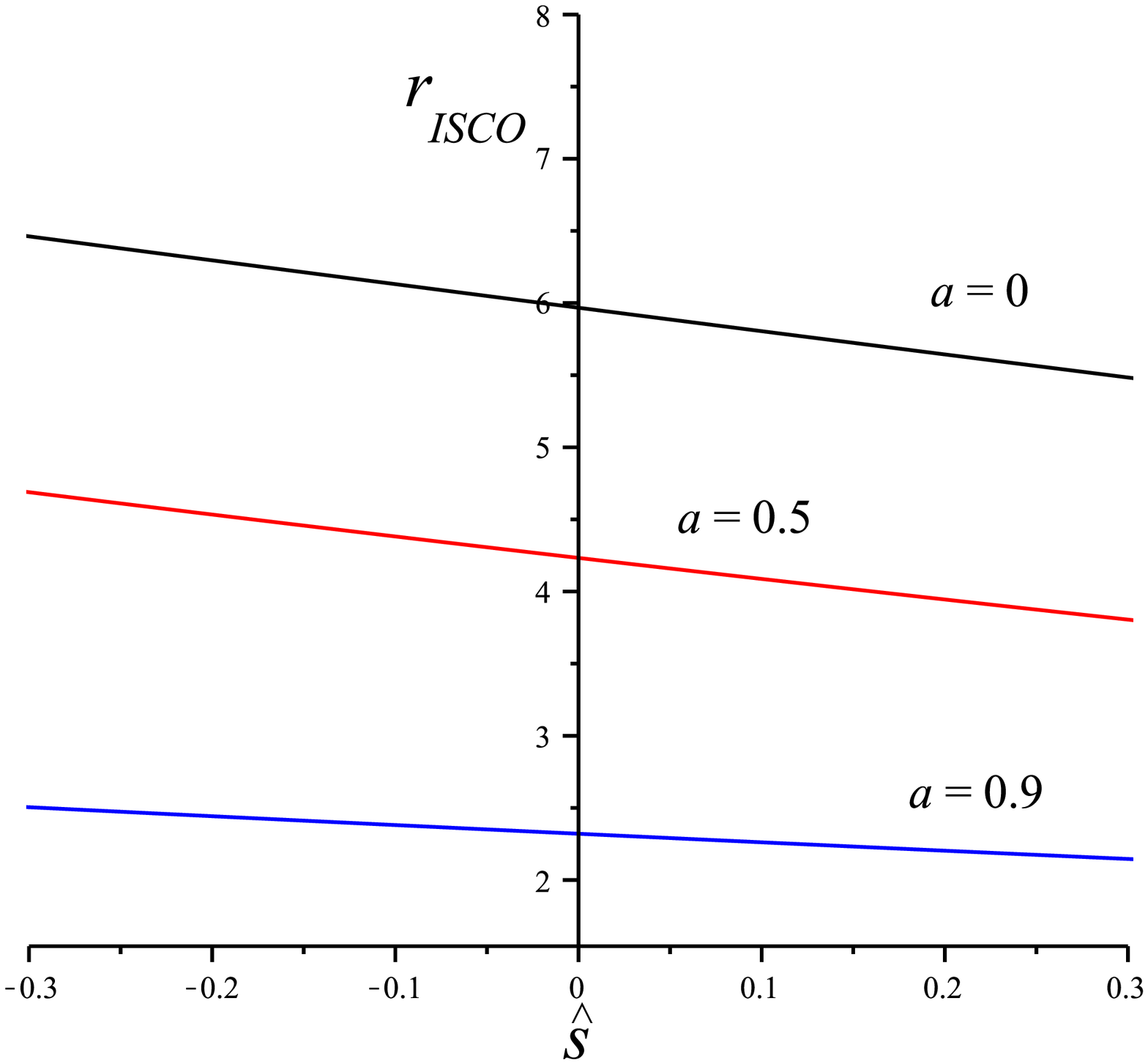}&\quad
\includegraphics[scale=0.3]{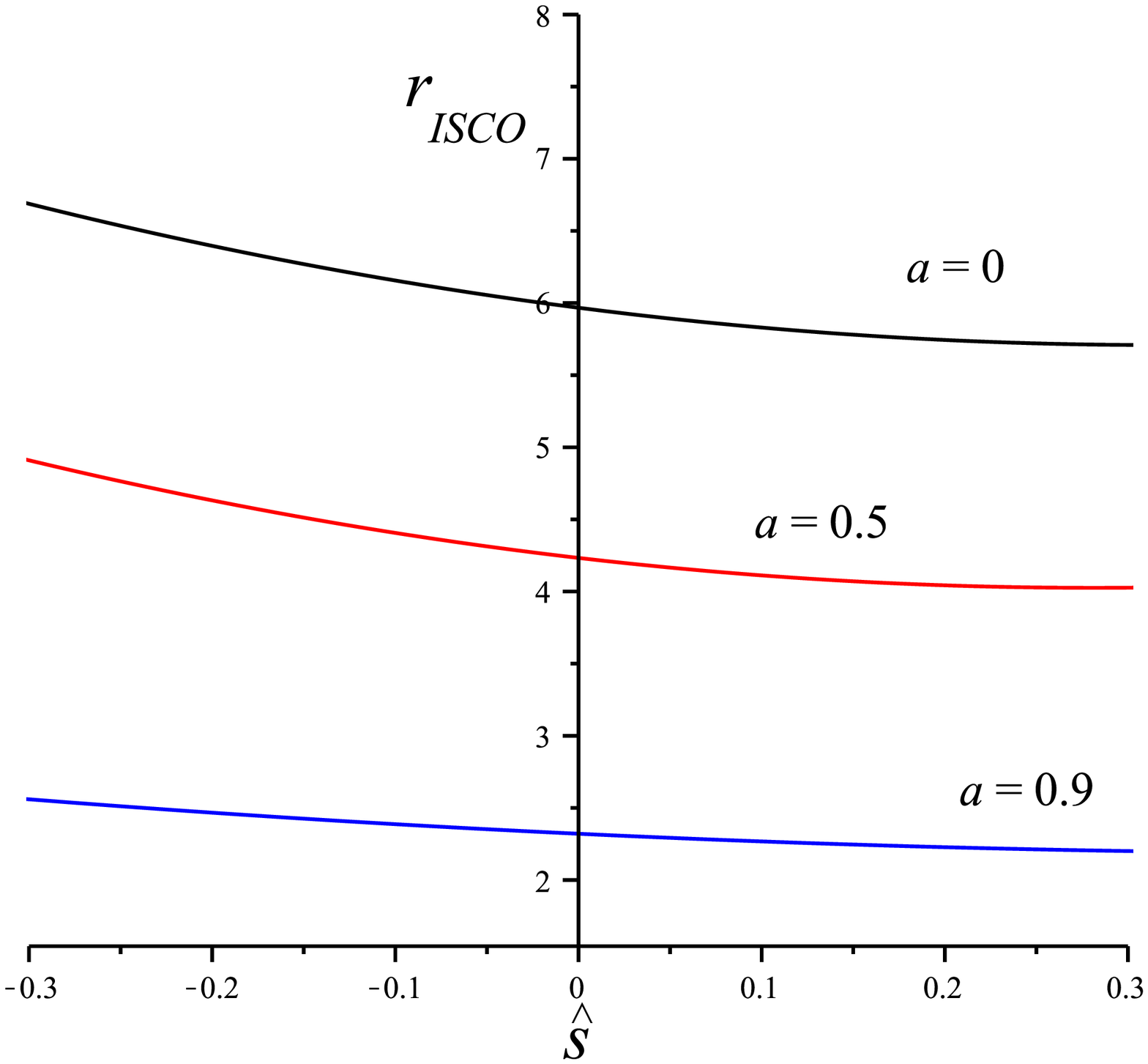}\\[.4cm]
\quad\mbox{(a)}\quad &\quad \mbox{(b)}
\end{array}$\\
\end{center}
\caption{The radius of the ISCO as a function of the spin parameter is shown
  in the case of co-rotating orbits for different values of the black-hole
  dimensionless spin, $a/M=[0,0.5,0.9]$, and for (a) $C_Q=1$, as well as (b)
  $C_Q=6$.}
\label{fig:risco_vs_s}
\end{figure}

For a spinning particle the ISCO is modified as follows:
\beq
r_{\rm ISCO}=r_{{\rm ms}}^{\rm Kerr}+{\hat s}r_{{\rm ms}(1)}
+{\hat s}^2r_{{\rm ms}(2)}\,,
\eeq
where
\begin{align}
\frac{r_{{\rm ms}(1)}}{M}&=\mp\frac{2\sqrt{6}}{3}
+\frac29{\hat a}\pm\frac{17\sqrt{6}}{324}{\hat a}^2
+\frac{43}{486}{\hat a}^3+O({\hat a}^4)\,,\nonumber\\
\frac{r_{{\rm ms}(2)}}{M}&=-\frac{29}{72}
+\frac12C_Q\pm\frac{\sqrt{6}}{54}\left(\frac{23}{24}+C_Q\right){\hat a}
+\frac{1}{81}\left(\frac{451}{48}-4C_Q\right){\hat a}^2
\pm\frac{\sqrt{6}}{1296}\left(\frac{4559}{72}-37C_Q\right){\hat a}^3
+O({\hat a}^4)\,.
\end{align}
Of course, both the first and second order corrections to the ISCO can also be
straightforwardly computed in the strong field regime. However, the
corresponding expressions are quite long, so we prefer not to explicitly write
them down. For instance, in the co-rotating case, for ${\hat a}=0.5$, we find
$r_{{\rm ms}}^{\rm Kerr}/M\approx4.233$, $r_{{\rm ms}(1)}/M\approx-1.472$ and
$r_{{\rm ms}(2)}/M\approx0.194$ ($C_Q=1$), or $r_{{\rm
    ms}(2)}/M\approx2.674$ ($C_Q=6$). The resulting behavior of $r_{\rm ISCO}$
as a function of the spin parameter ${\hat s}$ is shown in
Fig.~\ref{fig:risco_vs_s}.

\begin{figure} 
\typeout{*** EPS/PDF figure 5}
\begin{center}
$\begin{array}{cc}
\includegraphics[scale=0.3]{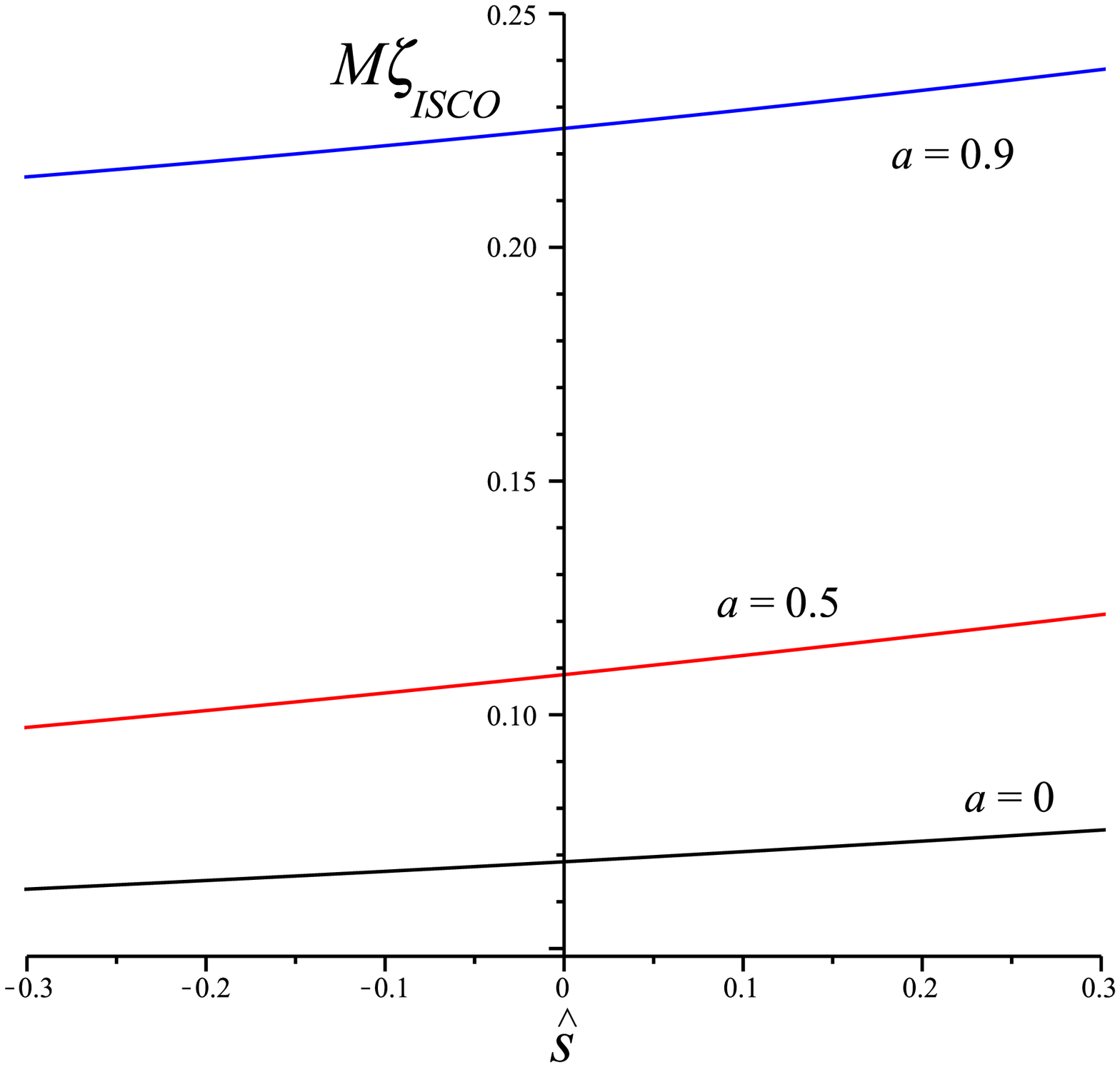}&\quad
\includegraphics[scale=0.3]{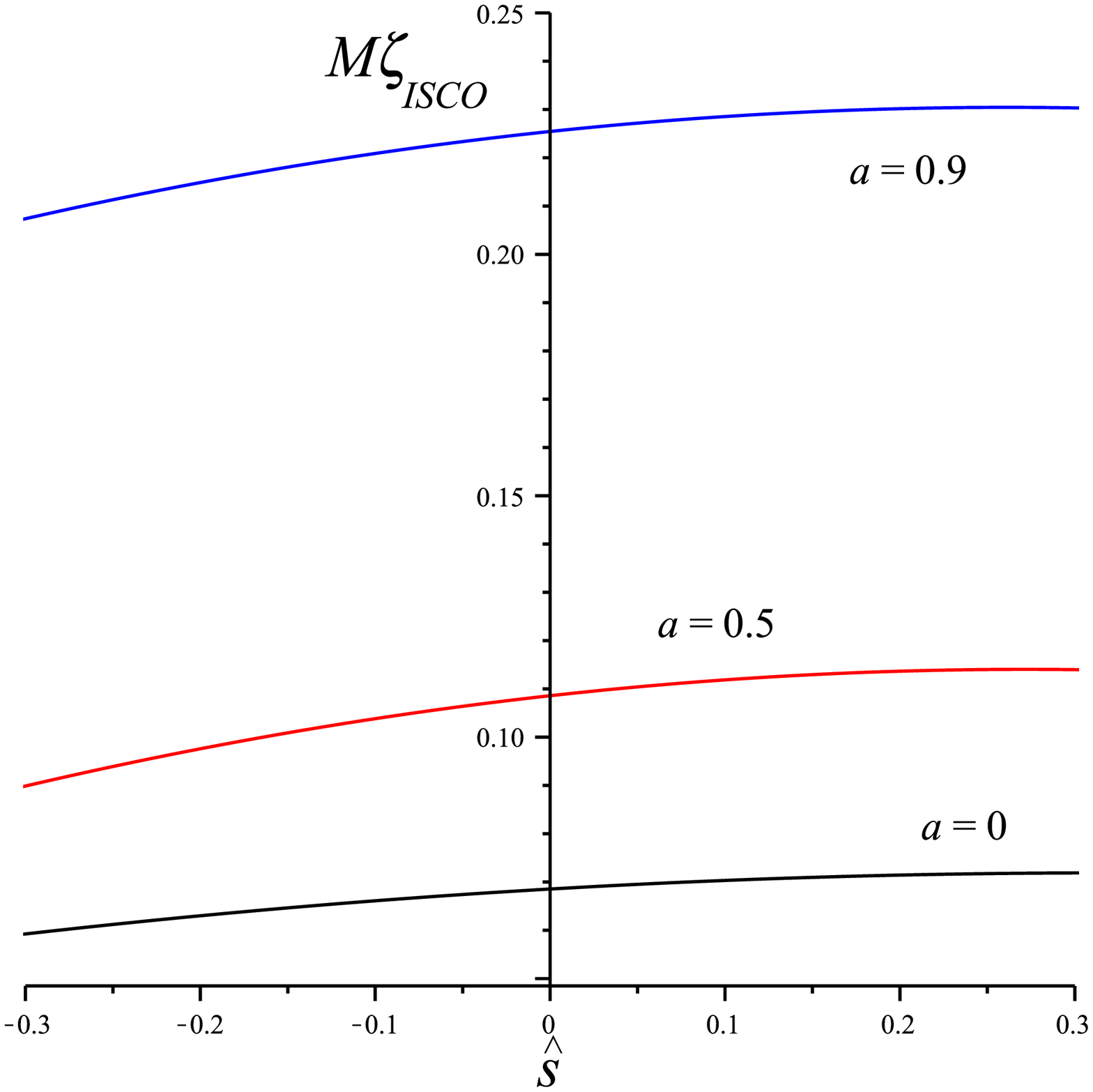}\\[.4cm]
\quad\mbox{(a)}\quad &\quad \mbox{(b)}
\end{array}$\\
\end{center}
\caption{The behavior of the ISCO frequency as a function of the spin
  parameter is shown in the case of co-rotating orbits for different values of
  the black-hole spin, $a/M=[0,0.5,0.9]$, and for (a) $C_Q=1$, as well as (b)
  $C_Q=6$.}
\label{fig:zetaisco_vs_s}
\end{figure}

\begin{figure} 
\typeout{*** EPS/PDF figure 6}
\begin{center}
$\begin{array}{cc}
\includegraphics[scale=0.3]{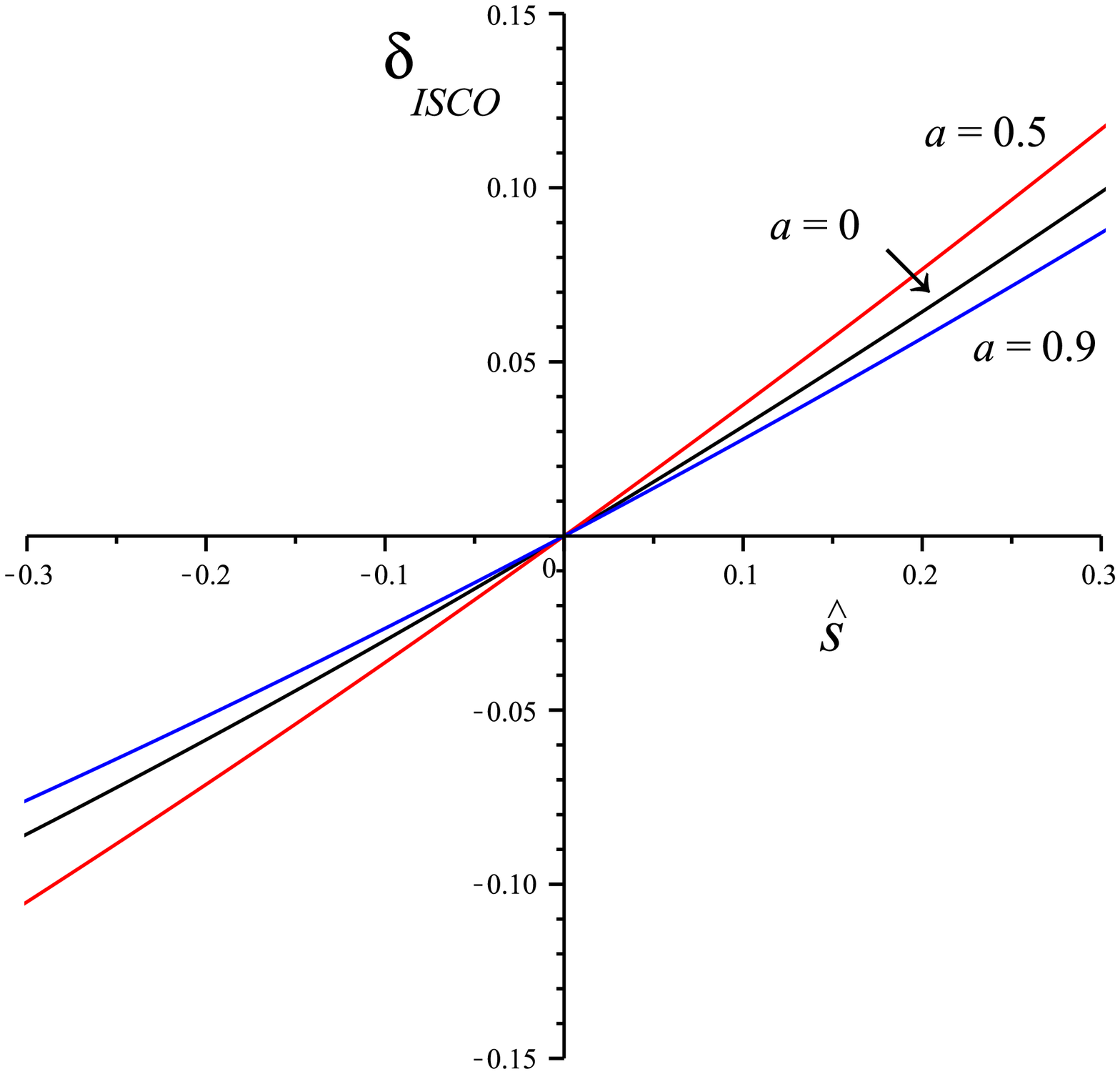}&\quad
\includegraphics[scale=0.3]{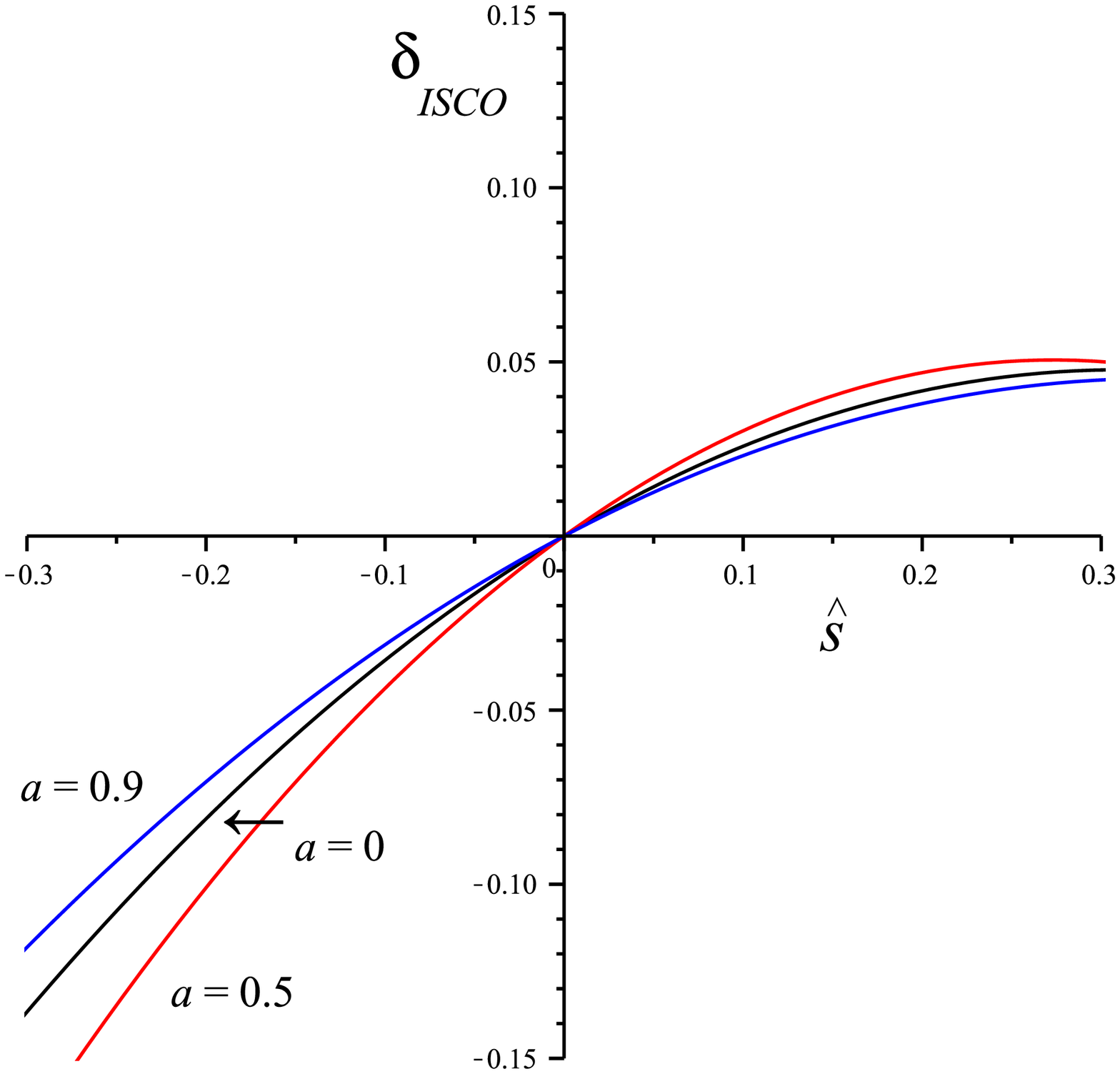}\\[.4cm]
\quad\mbox{(a)}\quad &\quad \mbox{(b)}
\end{array}$\\
\end{center}
\caption{The behavior of the fractional correction to the ISCO frequency as a
  function of the spin parameter is shown in the case of co-rotating orbits
  for different values of the black-hole dimensionless spin,
  $a/M=[0,0.5,0.9]$, and for (a) $C_Q=1$, as well as (b) $C_Q=6$. }
\label{fig:deltaisco_vs_s}
\end{figure}

The ISCO frequency of a spinning test object is then computed to be
\beq
M\zeta_{\rm ISCO}=\pm6^{-3/2}
\left\{1\pm\frac{1}{\sqrt{6}}\left(\frac{11}{6}{\hat a}
+\frac{3}{4}{\hat s}\right)
+\frac{59}{108}{\hat a}^2+\frac13{\hat a}{\hat s}
+\frac{1}{9}\left(\frac{97}{64}-C_Q\right){\hat s}^2
\right\}\,,
\eeq
leading to a fractional correction with respect to the spinless case
\beq
\delta_{\rm ISCO}\equiv\frac{\zeta_{\rm ISCO}}{\zeta_{\rm ISCO}^{\rm Kerr}}-1
=\pm\frac{\sqrt{6}}{8}{\hat s}+\frac{5}{48}{\hat a}{\hat s}
+\frac{1}{9}\left(\frac{97}{64}-C_Q\right){\hat s}^2
\,,
\eeq
whereas the energy and angular momentum at the ISCO read
\begin{align}
\frac{E_{\rm ISCO}}{m_0}&=\frac{2\sqrt{2}}{3}\mp
\frac{\sqrt{3}}{108}(2{\hat a}+{\hat s})
-\frac{2\sqrt{2}}{3}\left[\frac{5}{216}{\hat a}^2
+\frac1{54}{\hat a}{\hat s}
+\frac{1}{216}\left(\frac{15}{8}-C_Q\right){\hat s}^2\right]
\,,\nonumber\\
\frac{J_{\rm ISCO}}{m_0M}&=\pm2\sqrt{3}-\frac{\sqrt{2}}{3}(2{\hat a}-{\hat s})
\mp2\sqrt{3}\left[\frac{2}{27}{\hat a}^2
+\frac{11}{216}{\hat a}{\hat s}
+\frac{1}{24}\left(1-\frac{7}{9}C_Q\right){\hat s}^2\right]
\,.
\end{align}
The behavior of the ISCO frequency as a function of the spin parameter ${\hat
  s}$ is shown in Fig. \ref{fig:zetaisco_vs_s} for two typical values of
$C_Q$, i.e., $C_Q=1$ (black hole) and $C_Q=6$ (neutron star). The
corresponding curve is in general a parabola, which is concave up or down
depending on whether the sign of the coefficient of ${\hat s}^2$ is positive
or negative. For instance, for the
chosen values of the rotation parameter ${\hat a}=[0,0.5,0.9]$, the change of
concavity (from up to down) occurs at $C_Q\approx[1.516,1.487,1.455]$,
respectively. Finally, Fig. \ref{fig:deltaisco_vs_s} shows the behavior of the
fractional correction to the ISCO frequency as a function of the spin
parameter. Furthermore, restricting $\hat{s}$ to a given range of values
yields the uncertainty associated with the ISCO position, angular velocity and
shift. As an example, we list the three latter quantities below in Table
\ref{tab:1} for selected values of the rotational parameter ${\hat
  a}=[0,0.1,0.3,0.5,0.7,0.9]$, spin parameter ${\hat s}=[-0.1,0,0.1]$ and
polarizability parameter $C_Q=[1,6]$, in the corotating case.


%
%


%
%


%
%


%



%
\begin{table}
\centering
\caption{The numerical values for the modified ISCO position, angular velocity
  and shift are listed for selected values of ${\hat s}$ and ${\hat a}$ in the
  corotating case, for $C_Q=1$ (black hole), as well as $C_Q=6$ (neutron star).
}
\begin{tabular}{|c|ccc|ccc|ccc|}
\hline
\multicolumn{10}{|c|}{$C_Q=1$}\\
\hline
&\multicolumn{3}{|c|}{$r_{\rm ISCO}/M$}
&\multicolumn{3}{|c|}{$M\zeta_{\rm ISCO}$}
&\multicolumn{3}{|c|}{$\delta_{\rm ISCO}$}\\
\hline
\backslashbox{${\hat a}$}{\\ ${\hat s}$}&$-0.1$ &0 &0.1 &$-0.1$ &0 &0.1 
&$-0.1$ &0 &0.1\\
\hline
0 & 6.1643 & 6 & 5.8377 & 0.0660 &  0.0680 & 0.0702 
& $-0.0300$ &  0 &  0.0312 \\
0.1 & 5.8313& 5.6693& 5.5094& 0.0712&  0.0735& 0.0759& $-0.0311$ & 0& 0.0323\\
0.3 & 5.1351& 4.9786& 4.8248& 0.0847& 0.0877& 0.0907& $-0.0335$ & 0& 0.0348\\
0.5 & 4.3824& 4.2330& 4.0876& 0.1046& 0.1086& 0.1127& $-0.0365$ & 0& 0.0376\\
0.7 & 3.5259& 3.3931& 3.3002& 0.1384& 0.1439& 0.1475& $-0.0379$ & 0& 0.0250\\
0.9 & 2.4018& 2.3209& 2.2427& 0.2194& 0.2254& 0.2317& $-0.0267$ & 0& 0.0277\\
\hline
\multicolumn{10}{|c|}{$C_Q=6$}\\
\hline
&\multicolumn{3}{|c|}{$r_{\rm ISCO}/M$}
&\multicolumn{3}{|c|}{$M\zeta_{\rm ISCO}$}
&\multicolumn{3}{|c|}{$\delta_{\rm ISCO}$}\\
\hline
\backslashbox{${\hat a}$}{\\ ${\hat s}$}&$-0.1$ &0 &0.1 &$-0.1$ &0 &0.1 
&$-0.1$ &0 &0.1\\
\hline
0 & 6.1893 & 6 & 5.8627 & 0.0656 & 0.0680 & 0.0698 & -0.0356 & 0 &  0.0256 \\
0.1  & 5.8565& 5.6693& 5.5346& 0.0708&  0.0735& 0.0755& $-0.0370$ & 0& 0.0264\\
0.3  & 5.1604& 4.9786& 4.8501& 0.0841& 0.0877& 0.0901& $-0.0402$ & 0& 0.0281\\
0.5  & 4.4065& 4.2330& 4.1122& 0.1038& 0.1086& 0.1119& $-0.0438$ & 0& 0.0301\\
0.7  & 3.5438& 3.3931& 3.2815& 0.1375&  0.1439& 0.1486& $-0.0446$ & 0& 0.0330\\
0.9  & 2.4094& 2.3209& 2.2504& 0.2184&  0.2254& 0.2306& $-0.0313$ & 0& 0.0230\\
\hline
\end{tabular}
\label{tab:1}
\end{table}


\subsubsection{Quasi-circular orbits}

Let us finally construct the quadratic-in-spin solution to the MPD equations
corresponding to a quasi-circular orbit, in the perturbative sense. The
initial conditions are chosen so that the world line of the extended body has
the same starting point as the reference circular geodesic at radius $r=r_0$
for vanishing spin. We also require that the two world lines are initially
tangent.

The orbit can be parametrized in a Keplerian-like form as follows
\cite{dam-derue1,wex}
\begin{align}
\label{quasikeplgen}
\frac{2\pi}{P}(t-t_0)&= \ell_t -e_t \sin \ell_t\,,\nonumber\\
\qquad\quad\quad r&=a_r(1-e_r \cos \ell_r)\,, \nonumber\\
\qquad\quad\quad \theta &=\frac{\pi}{2}\,,\nonumber\\
\frac{2\pi}{\Phi}(\phi-\phi_0)&= 
2\arctan\left(\sqrt{\frac{1+e_\phi}{1-e_\phi}}\tan\frac{\ell_\phi}{2}\right)\,,
\end{align}
where $a_r$ is some ``semimajor axis'', $e_t$, $e_r$ and $e_\phi$ are three
different ``eccentricities'', which would coincide in the Newtonian theory,
while $P$ and $\Phi$ denote the periods of $t$ and $\phi$ motions,
respectively (with an abuse of notation for $P$, not to be confused with the
body's 4-momentum). The quantities $\ell_t$, $\ell_r$ and $\ell_\phi$ are
functions of the proper time parameter $\tau$ on the orbit. They are
conveniently expressed in terms of the dimensionless variable 
$\ell =\Omega_{\rm (ep)} \tau $, where
\beq
\label{omegaep}
\Omega_{\rm(ep)} 
=|\Omega_\pm|\left[1-\frac{6M}{r_0}
-\frac{3a^2}{r_0^2}\pm 8a\zeta_K\right]^{1/2}
\eeq
denotes the well-known epicyclic frequency governing the radial perturbations
of circular geodesics. The quantities $\ell_t$, $\ell_r$, $\ell_\phi$ and
$\ell$ play the role of eccentric anomalies. Note that, for geodesics, the
above quantities reduce to $a_r=r_0$, $e_t=e_r=e_\phi=0$, $P=2\pi\Gamma_\pm$,
$\Phi= 2\pi\Omega_\pm$ and $\ell_t=\ell_r=\ell_\phi=\ell$. For non-vanishing
spin parameter $\hat{s}$, the semimajor axis and the eccentricities for the
orbit of the extended body turn out to be
\beq
\label{solarquad}
a_r = r_0-{\hat s}R_{\hat s}-{\hat s}^2C_{1{\hat s}}\,,
\eeq
and
\begin{align}
\label{soleccquad}
e_t &=-\frac{\Omega_{\rm (ep)}}{\Gamma_\pm}{\hat s}\left[
T_{\hat s}+{\hat s}\left(D_{1{\hat s}}
+\frac{\Omega_{\rm (ep)}}{\Gamma_\pm}T_{\hat s}^2\right)
\right]\,, \nonumber\\
e_r &=-\frac{{\hat s}}{r_0}\left[R_{\hat s}+{\hat s}
\left(C_{1{\hat s}}+\frac{R_{\hat s}^2}{r_0}\right)
\right]\,, \\
e_\phi &=\frac{\Omega_{\rm (ep)}}{\Omega_\pm}{\hat s}\left[
\bar\zeta_\pm T_{\hat s}+{\hat s}\left(E_{1{\hat s}}
+\frac{\Omega_{\rm (ep)}}{\Omega_\pm}\bar\zeta_\pm^2T_{\hat s}^2\right)
\right]\,,\nonumber
\end{align}
respectively. As for the periods of $t$ and $\phi$ motions, they may be
written as
\begin{align}
\label{solper1quad}
P&=2\pi\frac{\Gamma_\pm}{\Omega_{\rm (ep)}}\left[
1\mp2\gamma_\pm\nu_\pm
\frac{\zeta_K}{\Omega_{\rm (ep)}}{\mathcal V}^{(r)}_{\hat s}{\hat s}
+\frac{D_{4{\hat s}}}{\Gamma_\pm}{\hat s}^2
\right]\,, \nonumber\\
\Phi&=2\pi\frac{\Omega_\pm}{\Omega_{\rm (ep)}}\left[
1\mp2\gamma_\pm\nu_\pm\frac{\zeta_K}{\Omega_{\rm (ep)}}
\frac{\bar\zeta_\pm}{\zeta_\pm}{\mathcal V}^{(r)}_{\hat s}{\hat s}
+\frac{E_{4{\hat s}}}{\Omega_\pm}{\hat s}^2
\right]\,.
\end{align}
There remains to display the three eccentric anomaly parameters, $\ell_t$,
$\ell_r$ and $\ell_\phi$, as functions of $\ell$:
\begin{align}
\ell_t&=\ell+\frac{{\hat s}^2}{\Gamma_\pm}[D_{2{\hat s}}\Omega_{\rm (ep)}
\sin2\ell+D_{3{\hat s}}\ell\cos \ell]\,,\nonumber\\
\ell_r&=\ell+\frac{{\hat s}}{R_{\hat s}\Omega_{\rm (ep)}}
[2C_{2{\hat s}}\Omega_{\rm (ep)}\sin\ell-C_{3{\hat s}}\ell]\,,\nonumber\\
\ell_\phi&=\ell+\frac{{\hat s}^2}{\Omega_\pm}\left[
\Omega_{\rm (ep)}\left(E_{2{\hat s}}
-\frac{1}{4}\Omega_{\rm (ep)}^2\bar\zeta_\pm^2T_{\hat s}^3\right)\sin2\ell
+E_{3{\hat s}}\ell\cos \ell\right]\,.
\end{align}
The various coefficients entering the above formulas are listed in Appendix
\ref{const}.
Notice that the parametrization~\eqref{quasikeplgen} of the orbit is different
from the quasi-Keplerian one, used in Refs.~\cite{dam-derue1,wex}, due to the
presence of different parameters representing the $t$, $r$ and $\phi$ motions
instead of a single eccentric anomaly. The two forms agree to first
order in spin only, as shown in Ref.~\cite{quadrupkerr1}.

At this stage, we can derive explicit expressions for the conserved energy and
angular momentum~\eqref{totalenergy}:
\beq
\label{costmoto1quad}
\tilde E\equiv\frac{E}{m_0}=\tilde E_\pm+{\hat s}\tilde E_{\hat s}
+{\hat s}^2\tilde E_{\hat s\hat s}\,, \qquad
\tilde J\equiv\frac{J}{m_0}=\tilde L_\pm+{\hat s}\tilde J_{\hat s}
+{\hat s}^2\tilde J_{\hat s\hat s}\,,
\eeq
with 
\beq
\label{costmoto2}
\tilde E_{\hat s}=M\Omega_\pm\left(\frac{M}{r_0}\mp a\zeta_K\right)\,, \qquad
\tilde E_{\hat s}-\zeta_\pm\tilde J_{\hat s}= \mp\frac{M\zeta_K}{\Gamma_\pm}\,,
\eeq
and
\begin{align}
\label{costmoto2quad}
& \tilde J_{\hat s\hat s}=-\sqrt{g_{\phi\phi}}
\gamma_\pm^2{\mathcal V}^{(r)}_{\hat s}\left[
\left(1-C_Q\right)M\Omega_{\rm (ep)}
+\frac{2}{\gamma_\pm\nu_\pm}{\mathcal V}^{(r)}_{\hat s}
\right]
+\frac{m_2}{m_0}\tilde L_\pm
\,,\nonumber\\
& \tilde E_{\hat s\hat s}-\zeta_\pm\tilde J_{\hat s\hat s}
=2N\gamma_\pm({\mathcal V}^{(r)}_{\hat s})^2
+\frac{m_2}{m_0}\frac1{\Gamma_\pm}
\,,
\end{align}
where $\tilde E_\pm$ and $\tilde L_\pm$ are the energy and the azimuthal
angular momentum per unit (bare) mass for a circular geodesic, as given by
Eqs.~\eqref{EandLgeocirc}, the parameter $m_2$ representing the spin
correction~\eqref{solmr} to the mass of the body [see also Eq.~\eqref{mcirc}].

In the weak field limit the previous expressions become
\begin{align}
-2{\widehat E}\equiv&-2(\tilde E-1)
=u_0\left(1-\frac34u_0-\frac{27}{8}u_0^2-\frac{675}{64}u_0^3\right)
+2u_0^{5/2}\left(1+\frac92u_0+\frac{135}{8}u_0^2\right){\hat a}\nonumber\\
& \qquad \qquad \quad
-2u_0^{5/2}\left(1+\frac32u_0+\frac{27}{8}u_0^2\right){\hat s}
-u_0^3\left(1+\frac{15}{2}u_0\right){\hat a}^2
+2u_0^3\left(1+\frac52u_0\right){\hat a}{\hat s}\nonumber\\
& \qquad \qquad \quad
+u_0^3\left[C_Q+6\left(1+\frac{5}{12}C_Q\right)u_0\right]{\hat s}^2
+O(u_0^5)
\,,\nonumber\\
{\widehat J}\equiv&\frac{\tilde J}{M}
=u_0^{-1/2}\left(1+\frac32u_0+\frac{27}{8}u_0^2
+\frac{135}{16}u_0^3+\frac{2835}{128}u_0^4\right)
-3u_0\left(1+\frac{5}{2}u_0+\frac{63}{8}u_0^2
+\frac{405}{16}u_0^3\right){\hat a}\nonumber\\
& \quad~
+\left(1-\frac12u_0+\frac38u_0^2+\frac{27}{16}u_0^3
+\frac{675}{128}u_0^4\right){\hat s}
+u_0^{3/2}\left(1+5u_0+\frac{189}{8}u_0^2+\frac{405}{4}u_0^3\right){\hat a}^2
-3u_0^{7/2}{\hat a}{\hat s}\nonumber\\
& \quad~
-21u_0^{5/2}\left[1+\frac{1}{42}C_Q+\frac1{14}\left(71+\frac{3}{2}C_Q\right)u_0
+\frac{15}{56}\left(111+\frac{3}{2}C_Q\right)u_0^2\right]{\hat s}^2
+O(u_0^5)\,, 
\end{align}
for the co-rotating case. 
We list below the orbital elements~\eqref{solarquad}--\eqref{solper1quad}
expressed in terms of the gauge invariant quantities ${\widehat E}$ and
${\widehat J}$, related by
\begin{align}
{\widehat J}&=(-2{\widehat E})^{-1/2}\left\{
1+\frac98(-2{\widehat E})+\frac{243}{128}(-2{\widehat E})^2
+\frac{5373}{1024}(-2{\widehat E})^3
-2(-2{\widehat E})^{3/2}\left[1+\frac94(-2{\widehat E})
+\frac{81}{8}(-2{\widehat E})^2\right]{\hat a}\right.\nonumber\\
&\left. \qquad \qquad\qquad
+(-2{\widehat E})^{1/2}\left[1-\frac32(-2{\widehat E})
-\frac94(-2{\widehat E})^2-\frac{81}{8}(-2{\widehat E})^3\right]{\hat s}
+\frac12(-2{\widehat E})^2
\left[1+\frac{69}{8}(-2{\widehat E})\right]{\hat a}^2\right.\nonumber\\
&\left. \qquad \qquad \qquad
+(-2{\widehat E})^2\left[1+\frac{45}{8}(-2{\widehat E})\right]{\hat a}{\hat s}
+\frac12(-2{\widehat E})^2
\left[C_Q-3\left(15-\frac{7}{8}C_Q\right)(-2{\widehat E})\right]{\hat s}^2
+O[(-2{\widehat E})^4]
\right\}\,,
\end{align}
so that we may write them with the help of a single parameter, e.g., the
energy parameter $(-2{\widehat E})\equiv X$. We find
\begin{align}
\frac{a_r}{M}&=X^{-1}\left(
1-\frac34X-\frac{63}{16}X^2-\frac{1215}{64}X^3-\frac{27135}{256}X^4\right)
+X^{1/2}\left(1+\frac{51}{8}X
+\frac{5751}{128}X^2\right)(2{\hat a}+{\hat s})\nonumber\\
&
-X\left(1+\frac{63}{4}X+183X^2\right){\hat a}^2
-X\left(1+\frac{79}{4}X+213X^2\right){\hat a}{\hat s}\nonumber\\
&
-\frac12X\left[C_Q-\left(36+\frac{5}{4}C_Q\right)X
-3\left(86-\frac{1}{2}C_Q\right)X^2\right]{\hat s}^2
+O(X^{7/2})
\,,\nonumber\\
e_t&=-6X^{5/2}\left\{
\left(1+\frac{63}{8}X\right){\hat s} 
-X^{1/2}\left(1+\frac{97}{4}X\right){\hat a}{\hat s} 
-\frac12X^{1/2}\left[C_Q+\left(1+\frac{25}{4}C_Q\right)X\right]{\hat s}^2 
+O(X^2)
\right\}
\,,\nonumber\\
e_r&=3X^{3/2}\left\{
\left(1+\frac{41}{8}X\right){\hat s} 
-X^{1/2}\left(1+\frac{33}{2}X\right){\hat a}{\hat s} 
-\frac12X^{1/2}\left[C_Q-\left(10-\frac{7}{2}C_Q\right)X\right]{\hat s}^2 
+O(X^2)
\right\}
\,,\nonumber\\
e_\phi&=6X^{3/2}\left\{
\left(1+\frac{41}{8}X\right){\hat s} 
-X^{1/2}\left(1+\frac{31}{2}X\right){\hat a}{\hat s} 
-\frac12X^{1/2}\left[C_Q-\left(4-\frac{7}{2}C_Q\right)X\right]{\hat s}^2 
+O(X^2)
\right\}
\,.
\end{align}
At last, the periods of $t$ and $\phi$ motions are given by
\begin{align}
\frac1M\frac{P}{2\pi}&=X^{-3/2}\left\{
1+\frac{15}{8}X+\frac{855}{128}X^2+\frac{41175}{1024}X^3
-12X^{5/2}\left(1+\frac{51}{4}X\right){\hat a}
-3X^{3/2}(1+6X+39X^2){\hat s}\right.\nonumber\\
&\left.
\qquad \qquad \, +11X^3{\hat a}^2
+3X^2\left(1+\frac{99}{8}X\right){\hat a}{\hat s}
+\frac32X^2\left[C_Q+\left(1+\frac{59}{8}C_Q\right)X\right]{\hat s}^2 
+O(X^4)
\right\}
\,,\nonumber\\
\frac{\Phi}{2\pi}&=
1+3X+\frac{63}{4}X^2+\frac{405}{4}X^3
-4X^{3/2}\left(1+\frac{93}{8}X\right){\hat a}
-6X^{3/2}\left(1+\frac{57}{8}X\right){\hat s}\nonumber\\
&
+\frac32X^2\left(1+\frac{73}{2}X\right){\hat a}^2
+6X^2\left(1+\frac{39}{2}X\right){\hat a}{\hat s}
+3X^2\left[C_Q+\frac12\left(13+11C_Q\right)X\right]{\hat s}^2 
+O(X^{7/2})
\,.
\end{align}
%


\section{Concluding remarks}

We have investigated finite-size effects on the motion of extended test bodies,
in the equatorial plane of a Kerr spacetime, within the framework of the
Mathisson-Papapetrou-Dixon model up to the quadrupolar order. In general, the
quadrupole tensor shares the same symmetries as the Riemann tensor and is
completely specified by two symmetric, trace-free spatial tensors, i.e.,
the mass quadrupole (electric) and the current quadrupole (magnetic) tensors,
whose role has been investigated in previous works
\cite{quadrupkerr1,quadrupkerrnum}. Here we have considered the rotational
deformation induced by a quadrupole tensor of the electric-type only, taken to
be proportional to the trace-free part of the square of the spin tensor, with a
constant proportionality parameter which may be regarded as the polarizability
of the object. This allows us to treat on an equal footing the cases of black
holes and neutron stars, so generalizing previous works.

The general features of equatorial motion have been discussed through the
analysis of the associated radial effective potentials. We have obtained their
generalization from the well-known case of a co/counter-rotating test monopole
particle in a Kerr spacetime to that of an extended test body with
spin-induced quadrupole moment. We have also evaluated the correction to the
ISCO due to spin and the corresponding frequency, which is an important
observable in gravitational-wave astronomy. The presence of spin corrections
introduce an uncertainty on the values of the corresponding quantities for
structureless particles. On the other hand, those features can be used
to determine whether the small object is endowed with a spin, by performing
an adequate parameter estimation in the context of gravitational-wave detection.

The dynamics of the system have been studied not only qualitatively, but also
quantitatively. In fact, neglecting terms in the MPD equations that are of
third order in spin or higher allowed us to solve the problem in a full
analytic way. Initial conditions have been chosen so that the tangent vector
to the orbit of the extended body be initially tangent to the 4-velocity of a
timelike spatially circular geodesic, taken as the reference trajectory. We
have obtained the ``perturbative'' solution to second order in spin in the
following two cases: (i) when the trajectory of the extended body remains
circular with spin-dependent frequency, (ii) when it deviates from circular
motion because of the combined effects of both the spin-curvature and
quadrupole-curvature couplings (i.e., when the orbit is \lq\lq
quasi-circular''). The tangent vector to the orbit and the unit timelike
vector aligned with the 4-momentum are in general distinct. However, there
exists a special value of the polarizability constant, which corresponds to
the black hole case, such that they are aligned not only initially, but all
along the (circular) trajectory of the extended body. This is no longer true
for neutron stars, an interesting fact which seems to have never been pointed
out before. For quasi-circular orbits, we have explicitly
written down the solution in a Keplerian-like form, by introducing the temporal,
radial and azimuthal eccentricities of the orbit, as well as the associated
periods and frequencies. We have also computed the spin-induced shift of the
conserved energy and angular momentum, in a gauge-invariant way. All orbital
elements have been expanded in the weak field and slow motion limit,
in a more suitable form to be compared with the existing post-Newtonian
literature.


\acknowledgments DB and AG acknowledge ICRANet and INFN for partial support.
GF thanks the CNR for his support through the Istituto per le Applicazioni del
Calcolo ``M. Picone''.


\appendix


\section{$\boldsymbol{1+3}$ decomposition of the quadrupole 
tensor} \label{appJ}

Let us consider an orthonormal frame adapted to an observer family
characterized by the 4-velocity $u$ (with normalization $u\cdot u=-1$), say
$e_0=u$ and $\{e_a\}$, $a=1,2,3$, a triad of three unit spatial vectors
orthogonal to $u$. We shall introduce the compact notation 
$X_{\alpha \beta\ldots }u^\alpha =X_{u\beta\ldots}$ for tensor contraction. In addition,
we shall denote by 
${}^*\!X_{\rho\sigma\ldots }=\frac12\eta^{\alpha\beta}{}_{\rho\sigma}X_{\alpha\beta\ldots }$
and  $X^*_{\ldots \mu\nu}=\frac12X_{\ldots \gamma\delta}\eta^{\gamma\delta}{}_{\mu\nu}$
the left and right dual of a tensor, respectively. The standard $1+3$
decomposition of the Riemann tensor in terms of its electric (spatial and
symmetric) part $E(u)$, its magnetic (spatial and tracefree) part $H(u)$, and
its mixed (spatial and symmetric) part $F(u)$, defined by
\beq
E(u)_{\alpha\beta}= R_{\alpha u \beta u}\,,\quad
H(u)_{\alpha\beta}= -R^*_{\alpha u \beta u}\,,\quad
F(u)_{\alpha\beta}= {}^*\!R^*_{\alpha u \beta u}\,,
\eeq
respectively, leads to the identification of the $20$ original independent
components: 6 in $E(u)$, 8 in $H(u)$ and 6 in $F(u)$.

Similarly, since the algebraic symmetries of the quadrupole
$J_{\alpha\beta\gamma\delta}$ are the same as for $R_{\alpha\beta\gamma\delta}$, one
can decompose the former quantity in terms of the associated tensors
\beq
{\mathcal Q}(u)_{\alpha\beta}= J_{\alpha u \beta u}\,,\quad
{\mathcal W}(u)_{\alpha\beta}= -J^*_{\alpha u \beta u}\,,\quad
{\mathcal M}(u)_{\alpha\beta}= {}^*\!\!J^*_{\alpha u \beta u}\,.
\eeq
In so doing we identify its electric (spatial and symmetric) part ${\mathcal
  Q}(u)$, with 6 independent components, its magnetic (spatial and tracefree)
part ${\mathcal W}(u)$, with 8 independent components, and its mixed (spatial
and symmetric) part ${\mathcal M}(u)$, with 6 independent components. However,
$J$ enters the MPD dynamics only in certain combinations, through
specific contractions with the Riemann tensor or its derivative. Hence, the
number of effective components needed is reduced by half, as shown in detail
in Refs.~\cite{quadrupschw,quadrupkerr1}. The proof requires the replacement of
the mixed part ${\mathcal M(u)}$ by a new tensor ${\mathcal X}(u)$ (with the
same symmetries), according to
\beq
{\mathcal M}(u)={\mathcal Q}(u)+{\mathcal X}(u)\,,
\eeq
as well as the decomposition of both ${\mathcal X}(u)$ and ${\mathcal W}(u)$
in terms of their STF and pure-trace parts,
\beq
{\mathcal X}(u)={\mathcal X}(u)^{\rm STF}
+\frac13 [{\rm Tr} {\mathcal X}(u)]P(u)\,,\quad
{\mathcal W}(u)={\mathcal W}(u)^{\rm STF}
+\frac13 [{\rm Tr} {\mathcal W}(u)]P(u)\,,
\eeq
where
$[P(u)]^\alpha_{\phantom{\alpha}\beta}=\delta^\alpha_{\phantom{\alpha}\beta}+u^\alpha
u_\beta$ denotes the projector to the hyperplane orthogonal to $u$. Inserting
the resulting expression for $J$ into the equations of motion then cancels the
contribution of ${\mathcal Q}(u)$, which yields the following ``effective"
representation of the quadrupole tensor (valid only in the context of the MPD
model):
\begin{align}
J^{\alpha\beta}{}_{\gamma\delta}=\eta(u)^{\alpha\beta\mu}
\eta(u)_{\gamma\delta}{}^\nu [{\mathcal X}(u)]^{\rm STF}_{\mu\nu}+
2u^{[\alpha}[{\mathcal W}(u)]^{\rm STF} {}^{\beta ]}{}_\sigma 
\eta(u)^\sigma{}_{\gamma\delta}
+2u_{[\gamma}[{\mathcal W}(u)]^{\rm STF} {}_{\delta ]}{}_\sigma 
\eta(u)^\sigma{}_{\alpha\beta}\,,
\end{align}
with $\eta(u)_{\alpha\beta\gamma}=u^\mu \eta_{\mu\alpha\beta\gamma}$ defining
the space 3-volume form (see section~\ref{MPD}). Summarizing,
in basis components, we can write
\begin{align}
J^{0a}{}_{0b}&= [{\mathcal X}(u)^{\rm STF}]^a{}_b \, , \nonumber\\
J^{0a}{}_{bc}&= [{\mathcal W}(u)^{\rm STF}]^a{}_{d}\eta(u)^d{}_{bc}=
[{\mathcal W}(u)^{\rm STF}]^{*(u)}{}^a{}_{bc} \, , \nonumber\\
J^{ab}{}_{cd}&= \eta(u)^{ab r}\eta(u)_{cd}{}^s [{\mathcal X}(u)^{\rm STF}]_{rs}=
[^{*(u)}[{\mathcal X}(u)^{\rm STF}] ^{*(u)}]^{ab}{}_{cd}\, .
\end{align}
For convenience, we actually use the notation:
\beq
{\mathcal X}(u)^{\rm STF}=\widetilde {\mathcal X}(u)\,,\qquad 
{\mathcal W}(u)^{\rm STF}=\widetilde {\mathcal W}(u)\,.
\eeq
%


\section{$\boldsymbol{1+3}$ decomposition of the MPD equations}

It is useful to perform a $1+3$ splitting with respect to $U$ of the MPD
equations (\ref{papcoreqs1})--(\ref{papcoreqs2}). A key observation is that
the force term on the right-hand side of the first equation~\eqref{papcoreqs1}
is not spatial for the comoving observer with 4-velocity $U$, since $F_{\rm
  (spin)}\cdot U=0$ whereas $F_{\rm (quad)}\cdot U\not=0$. Recalling that the
operator $P(U)^\alpha_{\phantom{\alpha}\beta} =
\delta^\alpha_{\phantom{\alpha}\beta} + U^\alpha U_\beta$ represents a projector
perpendicularly to $U$, we see that
 \begin{align}
 \label{papcoreqs1_bis}
 \frac{{\rm D}P^{\mu}}{\rmd \tau} 
 &= F^\mu_{\rm (spin)} +[P(U)F_{\rm (quad)} ]^\mu 
+ \frac16 U^\mu \, \, J^{\alpha \beta \gamma \delta}
\frac{{\rm D}}{\rmd \tau} R_{\alpha \beta \gamma \delta}\nonumber\\
 &\equiv F(U)^\mu_{\rm (tot)} + U^\mu \, \,\frac{{\rmd m_J}}{\rmd \tau} 
- \frac16 U^\mu R_{\alpha \beta \gamma \delta}
\frac{{\rm D}}{\rmd \tau}J^{\alpha \beta \gamma \delta}\,,
 \end{align}
 where the force
\beq
 F(U)^\mu_{\rm (tot)}\equiv F^\mu_{\rm (spin)} +[P(U)F_{\rm (quad)} ]^\mu 
=[P(U)(F_{\rm (spin)} +F_{\rm (quad)})]^\mu
\eeq
is orthogonal to $U$ and the mass correction $m_J$ has been defined in
Eq.~\eqref{mJdef}.
In a second stage, we get from Eq.~\eqref{papcoreqs1_bis} 
\begin{align}
\label{eq:b4}
 \frac{{\rm D}P^{\mu}}{\rmd \tau} - U^\mu \, \,\frac{{\rmd m_J}}{\rmd \tau} 
 = F(U)^\mu_{\rm (tot)} 
- \frac16 U^\mu R_{\alpha \beta \gamma \delta}
\frac{{\rm D}}{\rmd \tau}J^{\alpha \beta \gamma \delta} \,,
\end{align}
or, equivalently,
\begin{align}
\label{eq:b5}
& U_\mu\frac{{\rm D}P^{\mu}}{\rmd \tau} 
+\frac{{\rmd m_J}}{\rmd \tau} =  
\frac16 R_{\alpha \beta \gamma \delta}
\frac{{\rm D}}{\rmd \tau}J^{\alpha \beta \gamma \delta}\,, \nonumber\\
& P(U)\left[\frac{{\rm D}P^{\mu}}{\rmd \tau}\right]  = F(U)^\mu_{\rm (tot)} \,,
\end{align}
which follows from projecting Eq.~\eqref{eq:b4} along $U$ and perpendicularly
to $U$.


\section{ZAMO relevant quantities}
\label{appzamos}

We list below the non-vanishing components of the electric and magnetic parts
of the Riemann tensor, as well as the relevant kinematical quantities as
measured by ZAMOs and evaluated in the equatorial plane.

The radial components of the acceleration and expansion vectors are given by
\beq
a(n)^{\hat r}=\frac{M}{r^2\sqrt{\Delta}}
\frac{(r^2+a^2)^2-4a^2Mr}{r^3+a^2r+2a^2M}\,,\quad
\theta_\phi(n)^{\hat r}=-\frac{aM(3r^2+a^2)}{r^2(r^3+a^2r+2a^2M)}\,,
\eeq
respectively. The expressions for the radial components of the curvature
vector are
\begin{align}
\kappa(r,n)^{\hat r}&=\frac{Mr-a^2}{r^2\sqrt{\Delta}}\,,\qquad
\kappa(\theta,n)^{\hat r}=-\frac{\sqrt{\Delta}}{r^2}\,,\nonumber\\
k_{\rm (Lie)}&=-\frac{(r^3-a^2M)\sqrt{\Delta}}{r^2(r^3+a^2r+2a^2M)}\,.
\end{align}
Finally, the nontrivial components of the electric and magnetic
parts of the Riemann tensor with respect to ZAMOs read
\begin{align}
\label{E_H}
E_{\hat r \hat r}&= 
-\frac{M(2 r^4+5 r^2 a^2-2 a^2 M r+3 a^4)}{r^4 (r^3+a^2r+2 a^2 M)}\,, 
\quad
E_{\hat \theta \hat \theta}=-E_{\hat \phi \hat \phi}- E_{\hat r \hat r}\,,\quad
E_{\hat \phi \hat \phi}=\frac{M}{r^3}\,,\nonumber\\
H_{\hat r \hat \theta}&=  
-\frac{3 M a (r^2+a^2) \sqrt{\Delta}}{r^4 (r^3+a^2 r+2 a^2 M)}\,.
\end{align}

In the limit of vanishing rotation parameter ($a\to0$), the previous
quantities reduce to
\beq
a(n)^{\hat r}=\frac{M}{Nr^2}\,,\quad
\theta_\phi(n)^{\hat r}=0\,,\quad
k_{\rm (Lie)}=-\frac{N}{r}\,, \quad
E_{\hat r \hat r}=-\frac{2M}{r^3}\,,\quad
H_{\hat r \hat \theta}=0\,,
\eeq
with $N=\sqrt{1-{2M}/{r}}$.


\section{Frame components of both spin and quadrupole terms}
\label{spinandquadframecompts}

We list below the explicit expressions for the components of both spin and
quadrupole terms with respect to the frame adapted to $u$.

The spin force is given by Eq.~\eqref{fspinframeu} with
\begin{align}
\label{fspinframeu2}
F_{\rm (spin)}^1&=s\gamma\gamma_u\left\{
\nu_u\cos2\alpha_u E_{\hat r\hat r}
+[\nu\cos(\alpha_u-\alpha)+\nu_u\cos^2\alpha_u]E_{\hat \theta\hat \theta}
+\sin\alpha_u[1+\nu\nu_u\cos(\alpha_u-\alpha)]H_{\hat r\hat \theta}
\right\}\,,\nonumber\\
F_{\rm (spin)}^2&=-{\rm sgn}(\nu_u)s\gamma\gamma_u^2\left\{
\nu_u[\sin2\alpha_u-\nu\nu_u\sin(\alpha_u+\alpha)]E_{\hat r\hat r}
+[\nu\sin(\alpha_u-\alpha)
+\nu_u\cos\alpha_u(\sin\alpha_u-\nu\nu_u\sin\alpha)]
E_{\hat \theta\hat \theta}\right.\nonumber\\
&\left. \qquad \qquad \qquad \quad\,
-[\cos\alpha_u(1+\nu\nu_u\cos(\alpha_u-\alpha))-2\nu\nu_u\cos\alpha)]
H_{\hat r\hat \theta} \right\}\,,
\end{align}
the remaining component $F_{\rm (spin)}^0$ following from the condition
$F_{\rm (spin)}\cdot U=0$, i.e.,
\begin{align}
\gamma_u(1-\nu\nu_u\cos(\alpha_u-\alpha))F_{\rm (spin)}^0
=F_{\rm (spin)}^1\nu\sin(\alpha_u-\alpha)
+{\rm sgn}(\nu_u)F_{\rm (spin)}^2\gamma_u(-\nu_u+\nu\cos(\alpha_u-\alpha))\,.
\end{align}
On the other hand, the spin quantity $D_{\rm (spin)}$ takes the form of
Eq.~\eqref{dspinframeu}, with
\beq
\label{dspinframeu3}
{\mathcal E}_{\rm (spin)}(u)=m\gamma\left[\nu\sin(\alpha_u-\alpha)\omega^{1}
+{\rm sgn}(\nu_u)\gamma_u(\nu\cos(\alpha_u-\alpha)-\nu_u)\omega^{2}\right]\,.
\eeq

Concerning the quadrupole contributions, one gets for the components of the
force~\eqref{fspinframeu}:
\begin{align}
\label{fquadframeu3}
F_{\rm (quad)}^1&=
-\frac{1}{4}\frac{s^2}{m}C_Q\left\{\gamma_u^2\nu_u\left[
b_1\nu_u\sin3\alpha_u+2(b_4\cos2\alpha_u+b_5)\right]
-\left[(b_1-2b_2)\gamma_u^2-b_1+\frac23(b_2+b_3)\right]\sin\alpha_u
\right\}
\,,\nonumber\\
F_{\rm (quad)}^2&=
-{\rm sgn}(\nu_u)\frac{1}{4}\frac{s^2}{m}C_Q\gamma_u
\bigg\{b_1\gamma_u^2\nu_u^2\cos3\alpha_u
+2\left[(a_1-b_4)\gamma_u^2-a_1\right]\nu_u\sin2\alpha_u \nonumber \\
& \qquad \qquad \qquad \qquad \quad~\,
+\left[(b_1+2c_3)\gamma_u^2-b_1-\frac23(c_1+c_2)\right]\cos\alpha_u
\bigg\}
\,,
\end{align}
with
\begin{align}
a_1&=(E_{\hat \theta\hat \theta}+2E_{\hat r\hat r})\theta_{\hat\phi}(n)^{\hat r}
+H_{\hat r\hat \theta}a(n)^{\hat r}
\,,\nonumber\\
a_2&=(2E_{\hat \theta\hat \theta}+E_{\hat r\hat r})a(n)^{\hat r}
-H_{\hat r\hat \theta}\theta_{\hat\phi}(n)^{\hat r}
\,,\nonumber\\
b_1&=-2(E_{\hat \theta\hat \theta}+2E_{\hat r\hat r})k_{\rm (Lie)}
+(2E_{\hat \theta\hat \theta}+E_{\hat r\hat r})a(n)^{\hat r}
-4H_{\hat r\hat \theta}\theta_{\hat\phi}(n)^{\hat r}
+\frac12\partial_{\hat r}E_{\hat \theta\hat \theta}
\,,\nonumber\\
b_2&=-(E_{\hat \theta\hat \theta}+2E_{\hat r\hat r})k_{\rm (Lie)}
+2H_{\hat r\hat \theta}\theta_{\hat\phi}(n)^{\hat r}
-\frac32\partial_{\hat r}E_{\hat \theta\hat \theta}
\,,\nonumber\\
b_3&=-2(E_{\hat \theta\hat \theta}+2E_{\hat r\hat r})k_{\rm (Lie)}
-2H_{\hat r\hat \theta}\theta_{\hat\phi}(n)^{\hat r}
\,,\nonumber\\
b_4&=H_{\hat r\hat \theta}k_{\rm (Lie)}
-3E_{\hat \theta\hat \theta}\theta_{\hat\phi}(n)^{\hat r}
+\partial_{\hat r}H_{\hat r\hat \theta}
\,,\nonumber\\
b_5&=H_{\hat r\hat \theta}k_{\rm (Lie)}
-(E_{\hat \theta\hat \theta}+2E_{\hat r\hat r})\theta_{\hat\phi}(n)^{\hat r}
-\partial_{\hat r}H_{\hat r\hat \theta}
\,,
\end{align}
as well as $c_1=c_3+2a_2$, $c_2=c_3-b_2$ and $c_3=[7b_2+4a_2+6(b_1-2b_3)]/5$,
whereas $F_{\rm (quad)}^0$ may be obtained by requiring that the coordinate
component $F_{{\rm (quad)}\,t}=0$ vanishes, which implies
\begin{align}
0=\gamma_u(F_{\rm (quad)}^0+{\rm sgn}(\nu_u)\nu_u F_{\rm (quad)}^2)
+\frac{\sqrt{g_{\phi\phi}}N^{\phi}}{N}\left[
F_{\rm (quad)}^1\cos\alpha_u
-\gamma_u\sin\alpha_u(\nu_u F_{\rm (quad)}^0+{\rm sgn}(\nu_u)F_{\rm (quad)}^2)
\right]\,.
\end{align}
Finally, the torque tensor $D^{\mu \nu}_{\rm (quad)}$, whose structure is
displayed in
Eqs.~\eqref{dspinframeu}--\eqref{dspinframeu2}, is such that
\begin{align}
\label{quadtorquecompts}
{\mathcal E}_{\rm (quad)}{}_1 &=\frac{s^2}{m}C_Q\gamma_u\cos\alpha_u\left[
\nu_u\sin\alpha_u(2E_{\hat r\hat r}+E_{\hat \theta\hat \theta})-H_{\hat r\hat \theta}
\right]\,,\nonumber\\
{\mathcal E}_{\rm (quad)}{}_2 &= \,{\rm sgn}(\nu_u)\frac{s^2}{m}C_Q\gamma^2_u\left[
\nu_u\cos2\alpha_uE_{\hat r\hat r}+\nu_u(1+\cos^2\alpha_u)
E_{\hat \theta\hat \theta}
+\sin\alpha_u(1+\nu_u^2)H_{\hat r\hat \theta}
\right]\,.
\end{align}

\subsection{The Schwarzschild limit}

We list below the corresponding expressions of both spin and quadrupole terms
in the limit of vanishing Kerr spin parameter.

The spin force~\eqref{fspinframeu} becomes
\begin{align}
F_{\rm (spin)}&=\frac{M}{r^3}\gamma\gamma_u s\bigg\{
3\gamma_u\nu_u\sin\alpha_u(\nu\cos\alpha-\nu_u\cos\alpha_u)u\nonumber\\
& \qquad \qquad ~\,
+\frac12[\nu_u(1-3\cos2\alpha_u)+2\nu\cos(\alpha_u-\alpha)]e_1\nonumber\\
& \qquad \qquad ~\,
-\frac12{\rm sgn}(\nu_u)\gamma_u[(2+\nu_u^2)\nu\sin(\alpha_u-\alpha)
+3\nu_u(\nu\nu_u\sin(\alpha_u+\alpha)-\sin2\alpha_u)]e_2
\bigg\}\,,
\end{align}
whereas the spin quantity $D_{\rm (spin)}$ is still given by Eq.
\eqref{dspinframeu} with components~\eqref{dspinframeu3}.

The quadrupole force~\eqref{fspinframeu} reduces to
\beq
F_{\rm (quad)}=F_{\rm (quad)}^1e_1
+F_{\rm (quad)}^2[-{\rm sgn}(\nu_u)\nu_u u+e_2]\,,
\eeq
where $\gamma_u[-{\rm sgn}(\nu_u)\nu_u u+e_2]$ represents a unitary and 
spacelike vector orthogonal to $n$ and
\begin{align}
F_{\rm (quad)}^1&=-\frac{3MN}{2r^4}\frac{s^2}{m}C_Q\gamma_u^2\sin\alpha_u
\left[1-\frac12\nu_u^2(1+5\cos2\alpha_u)\right]\,,\nonumber\\
F_{\rm (quad)}^2&=-{\rm sgn}(\nu_u)\frac{3MN}{2r^4}
\frac{s^2}{m}C_Q\gamma_u^2\cos\alpha_u
\left[1+\frac12\nu_u^2(3-5\cos2\alpha_u)\right]\,.
\end{align}
Finally, for the torque term as shown in Eq.~\eqref{dspinframeu}, we have 
\beq 
{\mathcal E}_{\rm (quad)}=
-\frac{3M}{2r^3}\frac{s^2}{m}C_Q\gamma_u\nu_u\sin2\alpha_u
[\omega^1-{\rm sgn}(\nu_u)\gamma_u\tan\alpha_u\omega^2]\,.
\eeq
%


\section{Quasi-circular orbits: coefficients}
\label{const}

We list below the various coefficients entering the quasi-circular orbit
solution~\eqref{quasikeplgen}:
\beq
R_{\hat s} = -\gamma_\pm\frac{\sqrt{\Delta}}{r_0\Omega_{\rm(ep)}}
{\mathcal V}^{(r)}_{\hat s}\,,\qquad	
T_{\hat s}=\pm2\frac{\gamma_\pm^2\nu_\pm}{N}\frac{\zeta_K}{\Omega_{\rm(ep)}^2}
{\mathcal V}^{(r)}_{\hat s}\,,
\eeq
with
\beq
{\mathcal V}^{(r)}_{\hat s}=-3\frac{M\sqrt{\Delta}}{r_0\Omega_{\rm(ep)}}
\frac{\gamma_\pm\zeta_\pm^2}{N^2}\left(\frac{a}{r_0}\mp r_0\zeta_K\right)\,,
\eeq
and
\begin{align} 
\label{solord2coeffs}
C_{1{\hat s}}&=
\gamma_\pm\frac{\sqrt{\Delta}}{r_0\Omega_{\rm(ep)}^2}\left\{
\nu_\pm(B_{1{\hat s}}\Omega_{\rm(ep)}-B_{3{\hat s}})
+\gamma_\pm({\mathcal V}^{(r)}_{\hat s})^2
\left(\kappa(r,n)^{\hat r}+k_{\rm (Lie)}\right)
\right\}\,, \nonumber\\
C_{2{\hat s}}&=
\gamma_\pm\frac{\sqrt{\Delta}}{2r_0\Omega_{\rm(ep)}}\left\{
\nu_\pm B_{2{\hat s}}
-\frac{\gamma_\pm}{2\Omega_{\rm(ep)}}\left(\kappa(r,n)^{\hat r}
+k_{\rm (Lie)}\right)({\mathcal V}^{(r)}_{\hat s})^2\right\}
\,, \nonumber\\
C_{3{\hat s}}&=
-\gamma_\pm\nu_\pm\frac{\sqrt{\Delta}}{r_0\Omega_{\rm(ep)}}B_{3{\hat s}}
\,, \nonumber\\
D_{1{\hat s}}&=
\mp2\gamma_\pm^2\nu_\pm^2\frac{\zeta_K}{N\Omega_{\rm(ep)}^3}
(B_{1{\hat s}}\Omega_{\rm(ep)}-2B_{3{\hat s}})
+\frac{{\mathcal V}^{(r)}_{\hat s}}{MN\Omega_{\rm(ep)}^2}\left\{
\gamma_\pm^2\nu_\pm M^2(\zeta_K^2-\Omega_{\rm(ep)}^2)\right. \nonumber\\
&\left.
+M\Omega_{\rm(ep)}\gamma_\pm^3{\mathcal V}^{(r)}_{\hat s}\left[
1\mp\frac{2\nu_\pm\zeta_K}{\Omega_{\rm(ep)}^2}\left(
(3\nu_\pm^2-4)k_{\rm (Lie)}\pm6\nu_\pm\zeta_K\right)\right]
+2\gamma_\pm^4\nu_\pm({\mathcal V}^{(r)}_{\hat s})^2
\right\}\,, \nonumber\\
D_{2{\hat s}}&=
\mp\gamma_\pm^2\nu_\pm^2\frac{\zeta_K}{2N\Omega_{\rm(ep)}^2}B_{2{\hat s}} \nonumber\\
&
-\frac{\gamma_\pm^3}{4N\Omega_{\rm(ep)}^2}({\mathcal V}^{(r)}_{\hat s})^2\left\{
\frac{\gamma_\pm\nu_\pm^2}{4M\Omega_{\rm(ep)}}{\mathcal V}^{(r)}_{\hat s}
+1\mp\frac{\nu_\pm\zeta_K}{\Omega_{\rm(ep)}^2}\left[
(3\nu_\pm^2-4)k_{\rm (Lie)}\pm6\nu_\pm\zeta_K\right]
\right\}
\,, \nonumber\\
D_{3{\hat s}}&=
\mp2\gamma_\pm^2\nu_\pm^2\frac{\zeta_K}{N\Omega_{\rm(ep)}^2}B_{3{\hat s}}
\,, \qquad \qquad \qquad \qquad
D_{4{\hat s}}=
-\Omega_{\rm(ep)}(D_{1{\hat s}}+2D_{2{\hat s}})-D_{3{\hat s}}
\,,\nonumber\\
E_{1{\hat s}}&=
\bar\zeta_\pm D_{1{\hat s}}
-\frac{\gamma_\pm}{\nu_\pm\sqrt{g_{\phi\phi}}\Omega_{\rm(ep)}}
({\mathcal V}^{(r)}_{\hat s})^2\,,
\qquad \qquad
E_{2{\hat s}}=
\bar\zeta_\pm D_{2{\hat s}}
+\frac{\gamma_\pm}{4\nu_\pm\sqrt{g_{\phi\phi}}\Omega_{\rm(ep)}}
({\mathcal V}^{(r)}_{\hat s})^2
\,,\nonumber\\
E_{3{\hat s}}&=
\bar\zeta_\pm D_{3{\hat s}}
\,,
\qquad \qquad \qquad \qquad \qquad \qquad \qquad
E_{4{\hat s}}=
\bar\zeta_\pm D_{4{\hat s}}
+\frac{\gamma_\pm}{2\nu_\pm\sqrt{g_{\phi\phi}}}({\mathcal V}^{(r)}_{\hat s})^2\,,
\end{align}
where
\begin{align} 
B_{1{\hat s}}&=
-2B_{2{\hat s}}-\frac{B_{3{\hat s}}}{\Omega_{\rm(ep)}}
-\frac{1}{\gamma_\pm\nu_\pm\Omega_{\rm(ep)}}
\left({\tilde F}^1_{\rm (quad)}\pm\zeta_K
\tilde{\mathcal E}_{\rm (quad)}^2\right)
+\frac{{\mathcal V}^{(r)}_{\hat s}}{\nu_\pm}\left[
\frac{\gamma_\pm^2}{\nu_\pm}\pm M\zeta_K
+\frac{4\gamma_\pm^2}{3M\Omega_{\rm(ep)}^2}(\nu_\pm k_{\rm (Lie)}\mp3\zeta_K)
({\mathcal V}^{(r)}_{\hat s})^2
\right]
\,,\nonumber\\
B_{2{\hat s}}&=
({\mathcal V}^{(r)}_{\hat s})^2\left\{
\frac{\gamma_\pm^2}{3\nu_\pm M\Omega_{\rm(ep)}^2}
(\nu_\pm k_{\rm (Lie)}\mp3\zeta_K){\mathcal V}^{(r)}_{\hat s}
+\frac{2\zeta_K^2}{\gamma_\pm\nu_\pm\Omega_{\rm(ep)}^3}
(k_{\rm (Lie)}\mp2\gamma_\pm^2\nu_\pm\zeta_K)
\right.\nonumber\\
&\left. \qquad \qquad
-\frac{\gamma_\pm}{2\nu_\pm}\frac{k_{\rm (Lie)}}{\Omega_{\rm(ep)}}
-\frac{\gamma_\pm^3}{6\nu_\pm\Omega_{\rm(ep)}^3}
\left[\nu_\pm^2\partial_{\hat r}E_{\hat \theta\hat \theta}
+2\nu_\pm\partial_{\hat r}H_{\hat r\hat \theta}-\partial_{\hat r}E_{\hat r\hat r}\right]
\right\}
\,,\nonumber\\
B_{3{\hat s}}&=-3\Omega_{\rm(ep)}B_{2{\hat s}}
+\frac{{\mathcal V}^{(r)}_{\hat s}}{\nu_\pm}\left\{
\frac{2\gamma_\pm^2}{M\Omega_{\rm(ep)}}
(\nu_\pm k_{\rm (Lie)}\mp3\zeta_K)({\mathcal V}^{(r)}_{\hat s})^2
-\frac32\gamma_\pm k_{\rm (Lie)}{\mathcal V}^{(r)}_{\hat s}
+\frac{M\zeta_K^2}{\Omega_{\rm(ep)}}(5\nu_\pm k_{\rm (Lie)}\mp2\zeta_K)
\right.\nonumber\\
&\left. \qquad \qquad \qquad \qquad \quad ~
-M\Omega_{\rm(ep)}(\nu_\pm k_{\rm (Lie)}\mp\zeta_K)
+\frac{M\gamma_\pm^2}{2\Omega_{\rm(ep)}}
\left[(\nu_\pm^2+1)\partial_{\hat r}H_{\hat r\hat \theta}
-\nu_\pm\partial_{\hat r}(E_{\hat r\hat r}-E_{\hat \theta\hat \theta})\right]
\right\}
\,,
\end{align}
and 
\begin{align}
\label{eqE5}
{\tilde F}^1_{\rm (quad)}&=\frac12C_Q\left\{
\gamma_\pm^2M^2\left[
(1+\nu_\pm^2)\partial_{\hat r}E_{\hat \theta\hat \theta}
+2\nu_\pm\partial_{\hat r}H_{\hat r\hat \theta}
\right]\right.\nonumber\\
&\left.
+2\frac{M\Omega_{\rm(ep)}\gamma_\pm{\mathcal V}^{(r)}_{\hat s}}{1+\nu_\pm^2}
(\nu_\pm k_{\rm (Lie)}\pm\zeta_K)(1+2\nu_\pm^2)\right.\nonumber\\
&\left.
-M^2\left[2E_{\hat \theta\hat \theta}+E_{\hat r\hat r}+\frac{E_{\hat r\hat r}
-E_{\hat \theta\hat \theta}}{1+\nu_\pm^2}\right]
(k_{\rm (Lie)}\mp2\gamma_\pm^2\nu_\pm \zeta_K)\nu_\pm^2
\right\}\,,\nonumber\\
\tilde{\mathcal E}_{\rm (quad)}^2&=
C_QM\Omega_{\rm(ep)}\gamma_\pm{\mathcal V}^{(r)}_{\hat s}\,.
\end{align}
%


\bibliography{quadrup_kerr_squad_15_v16_arx} 

\end{document}